\documentclass[12pt,preprint]{aastex}
\usepackage{graphicx}
\def\spose#1{\hbox to 0pt{#1\h\> \> SS}}
\def\simlt{\mathrel{\spose{\lower 3pt\hbox{$\mathchar"218$}}
     \raise 2.0pt\hbox{$\mathchar"13C$}}}
\def\simgt{\mathrel{\spose{\lower 3pt\hbox{$\mathchar"218$}}
     \raise 2.0pt\hbox{$\mathchar"13E$}}}

\def\kms{km s$^{-1}$}

\def\deg{$^\circ$}

\def\mic{{$\mu$m}}

\def\h2o{H$_2$O}

\def\teff{$T_{\rm eff}$}

\def\aple{$\mathrel{\hbox{\rlap{\hbox{\lower4pt\hbox{$\sim$}}}\hbox{$<$}}}$}

\def\apge{$\mathrel{\hbox{\rlap{\hbox{\lower4pt\hbox{$\sim$}}}\hbox{$>$}}}$}

\def\k350{K3--50A}

\def\uchii{UC\ion{H}{2}}

\def\he1{\ion{He}{1}}

\def\feiii{\ion{Fe}{3}}

\def\kriii{\ion{Kr}{3}}

\def\seiv{\ion{Se}{4}}

\slugcomment{}
\lefthead{Blum {\it et al.}}
\righthead{\uchii \ \k350}

\begin{document}
\title{High Angular Resolution Imaging Spectroscopy of the Galactic
  Ultra--Compact \ion{H}{2} Region K3--50A}

\author{Robert D. Blum\footnote{NOAO Gemini Science Center, 950 North
Cherry Avenue, Tucson, Arizona, 85719} }

\author{Peter J. McGregor\footnote{Research School of Astronomy and Astrophysics, The Australian National University,
Cotter Road, Weston Creek, ACT 2611, Australia.} }

\begin{abstract}

Gemini North adaptive optics imaging spectroscopy is presented for the
Galactic ultra--compact \ion{H}{2} (\uchii) region K3--50A. Data were
obtained in the $K-$band using the Near--infrared Integral Field
Spectrograph (NIFS) behind the facility adaptive optics module ALTAIR
in natural guide star mode. The NIFS data cube reveals a complex
spatial morphology across the 0.1 pc scale of the 3$''$ \uchii \ region. 
Comparison of the nebular emission to Cloudy ionization models shows that the central source must have an effective temperature between about 37000 K and 45000 K with preferred values near 40000 K. Evidence is presented for sharp density variations in the nebula which are interpreted as a clearing of material nearest the central source. High excitation lines of \feiii \ and \seiv \ show that the  ionization of the nebula clearly changes with distance from the central source. A double lobed kinematic signature ($\pm$ 25 \kms) is evident in the Br$\gamma$ line map which may be related to the larger scale ionized flow detailed in earlier investigations. This signature and the large scale flow are not co-alighned, but they may still be related. Though clearly resolved from the nebula, the central source itself remains buried, and the NIFS spectrum shows no evidence of photospheric lines. 

\end{abstract}

 {\it
Facilities:} \facility{Gemini:Gillett ()}

\section{Introduction}

\k350 \ has been the subject of many studies in the optical/near--infrared and radio regimes and most wavelengths in between. It is highly obscured at short wavelengths and bright at long wavelengths. The object was originally part of the planetary nebulae survey of \citet{k65}, but was later reclassified as an \ion{H}{2} region. \cite{pf74} observed \k350 \ at mid--infrared wavelengths and showed the optical nebula was offset from the radio source due to a strong gradient in the foreground extinction. \citet{ww77} determined precise offsets between the radio, infrared and optical sources which validate the \citet{pf74} model. 

\k350, also known as the Galactic radio source G70.3--1.6, has been extensively observed at long wavelengths. \citet{tm84} provided a high angular resolution map at 2 cm which resolved the ionized gas into a clumpy arc--like or shell--like structure on scales smaller than 1$''$. \citet{kcw94} included this object in their catalog of \uchii \ regions, describing it as a core--halo type while \citet{depree94} observed the larger nebula in the continuum and radio recombination line H76$\alpha$. The radio morphology is extended in two lobes to the NW and SE and \citet{depree94} found a velocity gradient along the lobes which led them to suggest the structure was an ionized outflow extending 10's of arcseconds from the center of the \uchii \ region.

\citet{how96} presented Br$\gamma$ and Br$\alpha$ emission--line maps at greater than arcsecond resolution and followed this work \citep{how97} with molecular line maps which they modeled as arising from a large scale molecular torus \citep[roughly perpendicular to the ionized outflow of][]{depree94}. 

More recent work has centered on near--infrared and mid--infrared wavelengths and higher angular resolution data.  \citet{hof04} observed the nebula in the $K-$band using speckle techniques to provide an order of magnitude higher angular resolution than previous near--infrared work, and \citet{oka03} used the Subaru 8m telescpe and mid--infrared imager COMICS to obtain \aple 0.5$''$ resolution images and spectra near 10 \mic.

In this paper, we present $K-$band integral--field spectroscopy of \k350 \ obtained on the Gemini North 8m telescope with the Near--infrared Integral Field Spectrograph. These data were taken in conjunction with the facility adaptive optics system and result in the most detailed data cube yet for the core of \k350. In an earlier paper \citep{bm08}, we presented similar data for the \uchii \ region G45.45$+$0.6. In that case, several massive stars were identified by photospheric lines and other objects showed signatures of buried massive young stellar objects. G45.45$+$0.6 clearly harbors a rich cluster. \citet{mess07} make similar conclusions based on their ESO VLT SINFONI integral--field spectrograph observations for the \uchii \ region in  the cluster [DBS2003]8. 
Previous investigators have suggested the same may be true of \k350: \citet{col91}, \citet{oka03}, and \citet{hof04}. Our data identify multiple continuum sources in \k350 and support the multiple star model, but we also show that the excitation in the central few arcseconds of \k350 \ is likely dominated by a single object (the unresolved continuum peak). 

In the following, the distance to the nebula is based on the work of  \citet{har75}. 
The \citet{har75} distance is derived from a compilation of velocities and a rotation
model; Harris' distance needs to be reduced due to a revised distance to
the Galactic center \citep[used a distance of 10 kpc]{har75}. Adopting a value of 8 kpc gives a reduction by a factor of 0.8 resulting in a distance to the \ion{H}{2} region of  about 7 kpc. 

\section{Observations}

Data were obtained with the Near--infrared Integral Field Spectrograph
(NIFS) at the Cassegrain focus of the Gemini North Fredrick C. Gillett
8--m telescope on Mauna Kea, Hawaii on the night of 20 July, 2006
(HST). NIFS was used with the facility adaptive optics (AO) module
ALTAIR
\footnote{http://www.gemini.edu/sciops/instruments/altair/altairIndex.html}
in natural guide star (NGS) mode.

NIFS is fully described by \citet{pm03}; see also \citet{bm08} for a
more detailed description of similar observations with NIFS as
described here. Briefly, NIFS slices an approximately three arc second by three arc second field into 29 spectral segments of 0.1$''$ width
(the ``slit'' width) and $\sim$ 3$''$ in length. The scale
along the spatial dimension is 0.043$''$ pix$^{-1}$. The resulting
``spaxels,'' or spatial pixels, are thus 0.043$''$ $\times$ 0.1$''$ in size, and each contains a
full spectrum covering one of the near--infrared bands.

In the present paper, $K-$band spectra are presented for the Galactic
\uchii \ region \k350. The AO guide star used for ALTAIR is located
10.2$''$ to the south and west of \k350 \ and has an $R$ magnitude of
12.7 according to the USNO catalog (source id: U1200\_14213688). There
was thin cirrus at the time of the observations, and the seeing (at
5000 \AA, corrected to zenith) reported during the observations by
ALTAIR was approximately 0.45$''$. The observations were obtained at
relatively high airmass (1.5), and ALTAIR was run at 500 Hz.

The observations consisted of a single coadded frame taken on source
and a second frame obtained on a nearby ($\sim$ 50$''$ West) blank
field. Each frame had a total exposure time of 600 seconds
(15$\times$40s) on source and 600 seconds on sky. NIFS was oriented with a position  (PA) of 70 degrees east of north to correspond approximately to the orientation of the large scale ionized flow from \k350 \citep{depree94}. This PA is aligned with the outflow direction.

The spectral resolving power of NIFS in the $K-$band is $\lambda /
\Delta\lambda =$ 5160 which results in a linear dispersion of 2.13
\AA/pixel. This dispersion, combined with the large format array gives
a full wavelength coverage at $K$ of about 4200 \AA \ accounting for
some minor truncation in the final data cube due to the systematic
shift of wavelength in each slitlet from the staircase design of the
image slicer.

\section{Data Reduction}

The line maps and spectra presented here were obtained using the
Gemini NIFS IRAF\footnote{IRAF is distributed by the National
Optical Astronomy Observatories, which are operated by the Association
of Universities for Research in Astronomy, Inc., under cooperative
agreement with the National Science Foundation.} data reduction
package (version 1.9).

The NIFS IRAF package allows for full reduction to the ``image cube''
stage where a final cube has a roughly 60$\times$62 pixel image plane
and a 2040 pixel spectral depth. First, raw images are prepared for
reduction by standard Gemini procedures that create the FITS image
variance and data quality extensions. Next, the data are sky
subtracted, flatfielded, rectified spatially, and wavelength
calibrated \citep[see][for a full description of these
  steps]{bm08}. The last two steps result in a uniform
spatial--spectral trace for each image slice row and a linear
wavelength scale. It is important to remember that the final IRAF data
cube resamples each slitlet to two 0.05$''$ pixels for convenience;
the angular resolution in this dimension is still 0.1$''$.

A lamp image of Ar and Xe lines was used to determine the dispersion
along each slice and as a function of the spatial dimension of each
slice. A quadratic polynomial was used in this case which
produced typical uncertainties in the position of a spectral line of
approximately $\pm$ 0.2 \AA \ ($\sim$ 1/10 pixel). Final spectra are
interpolated to a linear wavelength solution.

The spectra were next corrected for telluric absorption by division by
the spectrum of an A0 V star (HIP 102568). The spectrum of HIP 102568
was corrected for intrinsic Br$\gamma$ absorption by fitting a Voigt
profile to the telluric spectrum. Particular care was taken in the fit
in order to remove this feature from the telluric standard as
accurately as possible.

The final wavelength calibrated, spatially rectified, and telluric
corrected images were transformed into data cubes. The spatial pixel
scale was resampled to 0.05$''$ in the fine dimension and block
replicated to 0.05$''$ in the course dimension providing for a uniform
scale as described in the previous section. Line maps presented below
were extracted from these cubes using NFMAP. 

The zero point of the wavelength calibration is confirmed by the positions of OH airglow lines in the sky image. Analysis of these lines shows the zero point is accurate to $\pm$ 2.5 \kms. 
The calibration results in an observed blueshift of the nebular lines of approximately  3.8 \AA \ (1.8 pixels or $-$53 \kms). Lines are identified on a spectrum extracted from a 2.5$''$ diameter aperture centered on the NIFS FOV; see Figure~\ref{spec}. The spectrum is flux calibrated using the results of \citet[][see below]{how96}.
The line positions and relative strengths compared to Br$\gamma$ for this integrated spectrum are given
in Table~1.

\section{Results}

An image of the continuum near  21700 \AA \ is shown in
Figure~\ref{c217} along with the radio continuum image from \citet{depree94}. 
The continuum passband extracted was 9 pixels (19.2 \AA) wide.
The appearance of sources in our continuum image is similar to
the speckle $K-$band image presented by \citet{hof04} who report an
angular resolution of 0.11$''$ in their image derived from
observations obtaddiained under 1.1$''$ seeing. \citet{hof04} identify 10
point sources in their image, though some of these appear more
extended than others, and this is consistent with the continuum
image shown in Figure~\ref{c217}. The brightest source in
Figure~\ref{c217} is source \#1 of \citet{hof04}.  The size of the
most compact source in our image, source \#8 of \citet{hof04}, is
about 3.7 pixels in the fine sampling direction suggesting a FWHM of
\aple 0.19$''$. Source \#1 is more extended than this in the present
image ($\sim$ 4.9 pixels), but this is due to crowding with nearby
blended sources and the intense nebular emission surrounding
it. \citet{hof04} report Source \#1 to be ``point like.''  

The 10 sources identified by \citet{hof04} are clearly seen in
Figure~\ref{c217} with the possible exception of their source \#2
which is close to source \#1 and not obvious in our image.  
The \citet{hof04} sources are indicated by filled circles in Figure~\ref{c217}. An
additional point source (named ``8N'') appears near $-0.6, -1.2$ which
is indicated by a single contour in Figure~2 of \citet{hof04}. In
addition to these objects, there are other knots or clumps which might
contain embedded sources, or may be density enhancements in the
\ion{H}{2} region. These clumps appear to match well the morphology in
Figure~1 of \citet{hof04}. An exception is the ``X-like'' structure
centered on source \#1 described by \citet{hof04}. We see no evidence
of this feature in the lower contours of the nebular emission even
though similarly bright clumps from the \citet{hof04} image do show up
in our image. The mid--infrared images of \citet{oka03} also do not show this feature, though diffraction spikes from the telescope secondary are present in their images.

\citet{oka03} identified four bright sources in the core region of \k350 \ covered
by Figure~\ref{c217} in 9--13 \mic \ images with somewhat lower angular resolution. 
The three brightest, OKYM1--3, were unresolved
as individual point sources. OKYM3 was associated with source \#1 by
\citet{hof04}. The source labeled STHO1 in Figure~\ref{c217} is the peak of [\ion{Ne}{2}] emission in the
maps of \citet{oka03}. These authors suggest STHO1 is the location of
an embedded massive star. There is no obvious point source in our
continuum image or the $K-$band image of \citet{hof04} at this
position.

The diamonds in Figure~\ref{c217} indicate the mid--infrared
sources identified by \citet{oka03}. The offsets between these sources
(OKYM 1--4 and STHO1) are taken from \citet{oka03} and the postion of
OKYM~3 is assumed to coincide with Source \#1 \citep[also assumed to
  be the case by][]{hof04}. \citet{hof04} identified a $K-$band source
with each of these \citet{oka03} sources within the positional
uncertainties of the mid--infrared image. These sources are the obvious
ones in Figure~\ref{c217} near each of the OKYM sources.

\subsection{Line Emission Maps}

The continuum subtracted Br$\gamma$ emission is shown in
Figure~\ref{brgmap}. \citet{how96} present narrow--band Brackett
imaging. Their data are seeing limited and do not resolve any of the
structure shown in Figure~\ref{brgmap}.  \cite{how96} report a
Br$\gamma$ flux of 3.7$\times$10$^{-12}$ erg cm$^{-2}$ s$^{-1}$ in a
3.4$''$ aperture centered on the Br$\gamma$ peak in their images. This covers 
approximately the NIFS field of view.
Equating this flux to the total counts in an approximately 2.5$''$
aperture results in a peak intensity (one pixel) of 5.5$\times$10$^{-15}$
erg cm$^{-2}$ s$^{-1}$ pix$^{-1}$ or 2.2$\times$10$^{-12}$
erg cm$^{-2}$ s$^{-1}$ arcsec$^{-2}$. The peak Br$\gamma$ emission is to the SW of the
continuum peak, although there is a second peak nearly as strong and
very compact (point like, superposed on extended emission) 
centered on the continuum peak (source \#1). The \citet{hof04} sources  \#4--\#7 are also centered on local Br$\gamma$ peaks. The rest of the \citet{hof04} continuum sources are offset from nearby local Br$\gamma$ emission peaks.

\he1 \  emission follows the Br$\gamma$ emission closely. The left panel of Figure~\ref{ratio} shows the ratio of \ion{He}{1} to Br$\gamma$ for the \he1  $^3$P$^{\circ}$--$^3$S line at 21127 \AA. The ratio is remarkably uniform over most of the FOV and has a value of approximately 0.04 (uncorrected for extinction). This value, which would be larger in the absence of extinction, already suggests a hot star (s) with effective temperature \apge 40000 K \citep{ben99, han02}.  The stronger \he1 line, $^1$S--$^1$P$^{\circ}$ at 20587 \AA, is not uniform, however, in comparison to Br$\gamma$ (see the right panel of Figure~\ref{ratio} and discussion in \S~5.2.3). 

Figure~\ref{spec} shows that strong emission from multiply ionized species exist in the \uchii \ region. These include [\ion{Fe}{3}], [\ion{Kr}{3}], and [\ion{Se}{4}].  The first and last are the strongest lines apart from \he1 and Br$\gamma$. Figure~\ref{feiii} shows the line maps for two of these lines. The [\feiii] \ emission is clumpy and peaks on the source OKYM4 while the [\seiv] \ emission is more intense and somewhat more concentrated to the central ``rectangle'' or ``diamond''. The [\seiv] \ emission has a maximum at the continuum peak ($-$0.2, 0.6) which is unresolved (i.e. compact). The [\feiii] \ emission also has a local maximum at the continuum peak, but it appears to be affected by one or more bad pixels at the very center. These lines (including [\kriii]) appear to be common in \uchii; see \citet{han02} and \citet{bm08}.
The ionization potentials of \ion{Fe}{3}, \ion{Fe}{4}, \ion{Kr}{3}, \ion{Kr}{4}, \ion{Se}{3}, and \ion{Se}{4} are 30.7, 54.8, 37.0, 52.5, 30.8, and 42.9 eV, respectively. This suggests that the region with strong [\seiv] emission corresponds with \ion{Fe}{4} (for which there are no identified lines). The range of species seen is consistent with the presence of \ion{He}{1} recombination lines (\ion{He}{1} ionization potential $=$ 24.4 eV) and the lack of \ion{He}{2} recombination lines (\ion{He}{2} ionization potential $=$ 54.4 eV).

Molecular hydrogen emission is also detected (see Figure~\ref{spec} and Table~1). The strongest line is the v$=$1--0 S(1) line at 21218 \AA. A map of the line is shown in Figure~\ref{h2} with the contours of the Br$\gamma$ emission overlaid. The projection of the H$_2$ emission seems to follow the ionized emission  with the peak emission coincident with the Br$\gamma$ peak. There is a local maximum at the position of the continuum peak as well. The location and excitation of the H$_2$ emission are discussed further in \S~5.4.

\subsection{Gas Kinematics}

Figure~\ref{velmap} shows the measured Br$\gamma$ line center velocity (in \kms) for the entire FOV. The mean observed Br$\gamma$ velocity of $-$50 \kms \ has been subtracted to show relative velocity differences in this map. This mean corresponds to a velocity with respect to the Local Standard of Rest (LSR) $V_{\rm LSR}$ $=$  $-$62.5 $\pm$ 5 \kms. The velocities are overplotted with the Br$\gamma$ emission contours. Each spatial pixel in the Br$\gamma$ map was fit with a Gaussian profile over a restricted velocity range of $-$250 to 50  \kms. This choice was made so that the \he1 line at 21642 \AA \ (see Table~2) did not interfere with the continuum fit. The velocity map shown in Figure~\ref{velmap} was made with a modified version of the Euro3D program \citep{e3d} produced by one of us (PJM) and called NF3D. 

The Br$\gamma$
velocity map shows a $\sim$ $\pm$ 25 \kms \ range in velocity centroids. This
should be compared with our FWHM velocity resolution of $\sim$ 60 \kms. 
Two Br$\gamma$ emission clumps are apparent in Figure~\ref{velmap}. These are
separated by 1$''$ $-$ 2$''$ and oriented approximately at a position angle of 45 degrees. 
A single Gaussian provides an adequate fit to the Br$\gamma$ emission over the entire FOV. However, as we show in the following section, there is a broad blue wing to the emission near the continuum peak. In what follows, we call the single Gaussian fit to the bright component of the line ``the principal component.'' 
When two Gaussians are fit to
spaxels near the continuum peak, the effect is to redshift the
principal component by a further $\sim$ 5 \kms \ in this vicinity.

The continuum peak is not located near any point of symmetry with respect to the velocity distribution. The nearest possible point source to the apparent center of the velocity distribution (Figure~\ref{c217}) is STHO1 which was not detected in our continuum map. This source is only detected in the mid--infrared [\ion{Ne}{2}] map of \citet{oka03} and not in any continuum image presented here or by \citet{oka03} or \citet{hof04}. 
  
\subsection{The \ion{He}{1} Spectrum and a Broad Br$\gamma$ Component}

\begin{figure}
\plotone{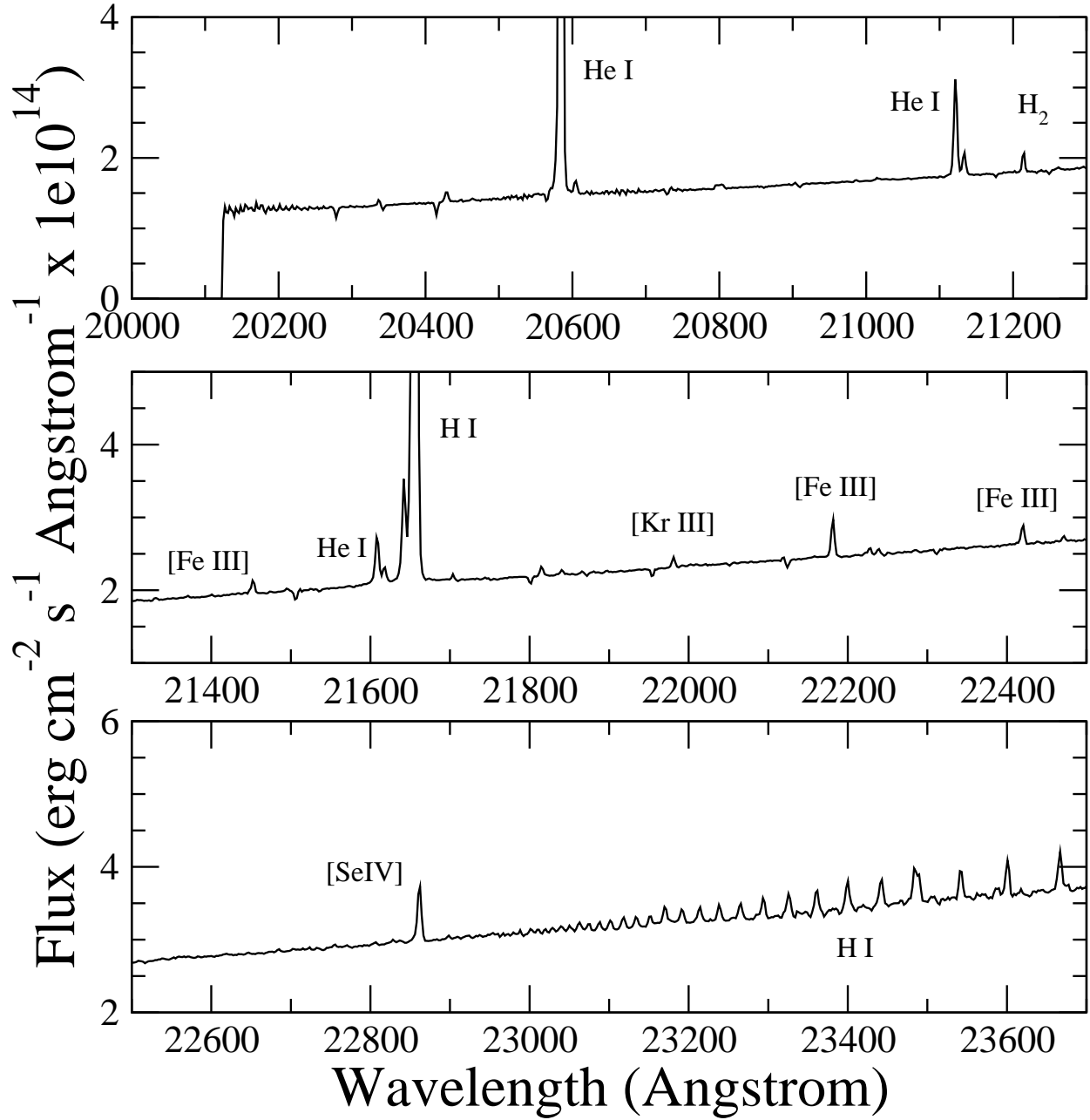}
\caption{$K-$band spectrum extracted from the central 2.5$''$ diameter of the NIFS FOV. Differences between the object and sky frames resulted in a slight over--subtraction. Line identifications and strengths relative to Br$\gamma$ are given in Table~1, and a number of the brighter lines are indicated in the figure. The \ion{H}{1} Pfund series is seen near 23000 \AA \ and redward.
\label{spec}}
\end{figure}

The \uchii \ region K3--50A exhibits a rich \he1 spectrum in the $K-$band. Table~2 lists the  \he1 lines detected and their properties. Using the on--line data base compiled by Peter van Hoof\footnote{http://www.pa.uky.edu/$\sim$peter/atomic/; the on--line database uses data from \citet{nist}.}; 13 lines are identified between 20000 and 22300 \AA. Other lines redward of 22300 \AA \ may be blended with the \ion{H}{1} Pfund series (Figure~\ref{spec}). Multiple \ion{He}{1} lines are present in the region around B$\gamma$. Figure~\ref{detail} shows  spectra extracted from individual
spaxels at a typical location $\sim$ 1.3$''$ SE of the continuum peak (Figure~\ref{spec}, left
panel) and a location $\sim$ 0.2$''$ SE of the continuum peak (Figure~\ref{spec}, right panel).
Five Gaussian functions have been fit simultaneously to these features using
the spectral analysis program Liner\footnote{Liner is provided by R. Pogge at the Ohio State University. We appreciate Dr. Pogge's support of the Liner package.}. Good fits with RMS residuals of $<$ 1$\%$
have typically been achieved by assuming that all five lines have the same
FWHM as Br$\gamma$ (Figure~\ref{spec}, left panel). 

Poorer fits with RMS residuals of $\sim$ 3$\%$ are obtained in the region within $\sim$
0.5$''$ of the continuum peak where the Br$\gamma$ line has a strong blue wing
(e.g., Figure~\ref{spec}, right panel). The Gaussian component fit to this wing has a
central wavelength (which is a fit parameter) that is close to that of the
\ion{He}{1} $4f-7d$ $^1$F$^{\circ}$--$^1$D, $^3$F$^{\circ}$--$^3$D 21655 \AA \ line (see Table~2). However, this line is
expected to be too weak to account for the blue wing near the continuum
peak. In fact, based on atomic data used in the Cloudy photoionization code
\citep{fer98}, the 21655 \AA \ \ion{He}{1} line is expected to be weaker than
the \ion{He}{1} $4f-7g$ $^1$F$^{\circ}$--$^1$G, $^3$F$^{\circ}$--$^3$G 21647 \AA \ line that is apparent in Figure~\ref{spec}.
Better fits are obtained for spaxels near the continuum peak if the FWHM of
the blue wing component is allowed to vary. RMS residuals of $<$ 1$\%$ are then
obtained with a Gaussian FWHM that is about twice as broad as the principal
Br$\gamma$ line. There is no apparent reason why the 21655 \AA \ \ion{He}{1} line should be
anomalously strong and have a width twice that of other \ion{He}{1} lines. We
therefore conclude that the blue wing on Br$\gamma$ is due to a mix of a second
kinematic component of Br$\gamma$ that dominates in a small region near the
continuum peak and weaker \ion{He}{1} emission that occurs over most of the FOV. It
is interesting that neither the peak emission nor the width of this second
kinematic component is centered on the continuum peak. Instead, this
emission forms an arc that is offset SE of the continuum peak by $<$ 0.5$''$. 

Whatever the source of the broad component, the contribution of the lines near 21600 \AA \ make a significant contribution to the integrated ``Br$\gamma$'' flux.  For the spaxel shown in Figure~\ref{detail}, the broad line produces 50$\%$ of the corresponding Br$\gamma$ flux. 

\section{Discussion}

\subsection{Source Positions \& Morphology}

\citet{ww77} measured the optical, near--infrared, and mid--infrared positions of \k350 \ with respect to the radio continuum \citep{cs77} using background stars as a reference.  The \citet{cs77} position is consistent with later radio positions including the high angular resolution study of \citet{tm84}, and thus, to our knowledge, the \citet{ww77} positions are still the best available. More recent studies have assumed the near and mid--infrared peaks are coincident with the radio continuum peak \citep{oka03, hof04}.  \citet{ww77} find that the 2.2 \mic \ and 10 \mic \  peaks are coincident, within the $\sim$ 1$''$ uncertainties, with the radio continuum peak. The optical and 1.6 \mic \ peaks are located to the south west of the radio peak, and this is consistent with an extinction gradient across \k350 \ as suggested by \citet{fp74} and \citet{ww77} and found also by \citet{how96}.

\citet{depree94} observed \k350 \ in the radio continuum (15 GHz) and in the H76$\alpha$ recombination line. The angular resolution of these data is about 1.5$''$. The continuum appears to be extended along an axis with PA (E from N) of 160 degrees and there is a strong velocity gradient from NW to SE along the extended continuum source \citep[see also][]{roel88}. This has led to the accepted model of an ionized outflow along the 160 degree axis. The NIFS continuum is shown superposed with the \citet{depree94} radio continuum image in Figure~\ref{c217}. The two images have been aligned assuming the radio and $K-$band continuum peaks are coincident. The large difference in scale and angular resolution are apparent. The overlay is instructive for the following discussion, but there are no obvious connections which can be drawn between features in the two maps.   

Molecular line studies have been used to argue that there is a rotating torus of material with major axis roughly coincident with and perpendicular to the radio source. Low angular resolution CO data and modeling of HCO$^+$ line emission suggests this material may be rotating about an axis parallel to the outflow axis. \citet{depree94} first drew attention to a possible link between their ionized outflow and the molecular gas traced by CO line emission \citep{pm91}, though they cautioned that the low angular resolution of the CO data (30$''$) and the marginal evidence for a velocity gradient along its major axis made any link between the two structures ``premature.'' Later, \citet{how97} used observations of HCO$^+$ to argue for a rotating torus roughly perpendicular to the ionized outflow. The peak lobes of HCO$^+$ emission are separated by 8$''$, so the scale of the rotating molecular torus is larger than the FOV covered by NIFS. The velocity difference measured between the peaks of the HCO$^+$ is $\sim$ $\pm$ 2 \kms, and the best fit to the velocity profile was found to be proportional to the square root of the radius.

Higher angular resolution observations were made by \citet{oka03} in the mid--infrared (near 10 \mic) and by \citet{hof04} in the $K-$band (2.2 \mic). The former observations were diffraction limited based on the 0.4$''$ FWHM obtained on standard stars, but the images of \k350 \ do not clearly separate individual objects owing to intense resolved structure/emission and blended point sources. \citet{oka03} fit various point sources (Figure~\ref{c217}) to their data. The \citet{hof04} images, on the other hand, were obtained with a speckle imaging technique and are of similar angular resolution as the data presented here. The higher angular resolution and lower background emission result in more identifiable point sources within the central few arcseconds of \k350. \citet{oka03} and \citet{hof04} conclude that \k350 \ contains multiple massive stars; the former through excitation measurements of mid-infrared ionized lines and the latter based on the detection of multiple point sources in the center of the \uchii \ region. 

The NIFS data confirm various aspects of the earlier work, principally the $K-$band morphology of the point sources and brighter clumps of emission which are dominated by Br$\gamma$. The $K-$band image of \citet{hof04}, the \citet{oka03} mid--infrared images, and our NIFS images all indicate an abrupt cutoff in emission to the north of the bright continuum peak. The gradient is extremely sharp and the emission drops precipitously in \aple 0.5$''$ towards the north of the \uchii \ region. The gradient in extinction was noted in earlier work \citep[e.g.,][]{ww77} but is seen very clearly in the $J, H,$ and $K$ images of \citet{oka03}. The center of the emission shifts systematically to the south. This geometry has been described as an opening cone with the continuum peak at the cone's apex \citep{hof04} and the opening toward the south--south east. 

The dense molecular torus \citep{how97} and other overlying material form the cone structure by obscuring ionized material to the north; see Figure~7 of \citet{how97}. The constraining effect of the torus may be responsible for helping to shape the ionized outflow \citep{depree94}. The very strong gradient in emission evident in Figure~\ref{c217} and the $K-$band image of \citet{hof04} suggest a possible obscuring disk on small scales, but no kinematic or morphological evidence exists to support this picture, and the radio and near--infrared peaks are offset from the center of the torus \citep[see Figure~14 of][]{oka03}.  If the embedded massive star associated with the continuum peak formed from the molecular torus, then it may have separated itself from this ``parental cloud.'' 

\subsection{Ionization of the Nebula}

\citet{oka03} presented ionic line measurements of collisionally excited \ion{Ar}{3}, \ion{S}{4}, and \ion{Ne}{2} at approximately 9, 11, and 13 \mic, respectively. Ratios of these lines suggest a relatively low excitation, typical of late O--type or early B-type stars. The radio continuum \citep{tm84, depree94}, however, requires more ionizing photons than a single late O--type star produces. \citet{oka03} thus concluded that multiple stars ionize \k350 (as discussed above for the NIFS images, we detect continuum sources, \#1, \#4--\#7 at least, which are centered on local peaks of the Br$\gamma$ emission).
\citet{oka01} made a similar analysis and conclusion for W51 IRS2, also a massive star forming region. But recently, \citet{bar08} identified the exciting star of W51 IRS2 directly through its near--infrared spectrum, and it is much hotter (approximately O3--O4). For G29.26$-$0.02, \citet{mor02} find a best fit nebular model which also has a cooler central source than the observed spectral type \citep{wat97}, though their acceptable models do cover a range of \teff \ which includes the observed spectral type. The nebular emission in G45.45$+$0.06 was consistent with the observed spectral types found by \citet{bm08}, but detailed modeling was not done.

A grid of ionization models has been run to analyze the near--infrared lines presented here in order to further explore the excitation of the \uchii \ region. Cloudy\footnote{Calculations were performed with version 07.02 of Cloudy, which was last described in detail by \citet{fer98}.} was run on a grid of input parameters which sample a range of ionizing flux and \teff \ of the central source as described by \citet{bm08}. A base line model grid was run with density 10$^{4}$ cm$^{-3}$ and the \citet{ck04} atmospheres (log (g) $=$ 5.0 and Z $=$ 0). The simulations were run until the temperature into the cloud reached 30 K. A model He atom was used with 10 energy levels to better resolve the level populations for the near--infrared lines. The ``\_ism'' abundance set available in Cloudy was used which has a He abundance by number relative to H of 0.10. In the following sections, we discuss how the NIFS observations constrain inputs to the models and also what can be inferred from a comparison of the models and observations.

\subsubsection{[\feiii] \ Lines}

The density used in the Cloudy runs was constrained by the observed ratio of [\feiii] lines. The ratio of [\feiii] 22184 \AA \ to 21457 \AA \ is sensitive to density \citep{lutz93, bau98}. The observed ratio (see Table~1) is 3.12 and corrected for A$_K$ of 2.17 \citep{how96} leads to a ratio of 2.8 and a density of approximately 10$^4$ cm$^{-3}$ according to the 7000 K model of \citet{lutz93} or the 10000 K model of \citet{bau98}. \citet{bau98} give densities for the ratio of [\feiii] 22427 \AA \ to 22184 \AA \ as well, but the observed ratio of 0.5 is on the flat part of their curve and gives a low density of \aple 10$^2$ cm$^{-3}$. The small reddening correction would push toward even lower density. All three lines are stronger toward the south in the FOV and weak near the continuum peak, thus the ratios do not constrain the density near the continuum peak. \citet{tm84} derived a density of 10$^5$ cm$^{-3}$ in the central few arcseconds from their 2 cm data; similar radio data were presented by \citet{kcw94} who find a density of 4$\times$10$^{4}$ cm$^{-3}$. We discuss models for a range of densities below (see \S5.2.2 and \S5.2.3 and Figure~\ref{density}).

The [\feiii] line ratios to Br$\gamma$  reported in Table~1 are similar to those reported by \citet{dp94} in Orion.  Though the emission is relatively strong in the \k350 \ spectrum, it may be consistent with most of the Fe being condensed onto dust grains as is the case for Orion; similar ratios and physical conditions in Orion are due to a gas phase abundance which is only a percent of the solar iron abundance \citep{dp94}. The distribution of \seiv \ and \feiii \ discussed above suggests some of the gas phase iron in \k350 \ must be in \ion{Fe}{4}, but there are no \ion{Fe}{4} lines to estimate the abundance in that species (the upper levels of observable lines are generally too high to be excited in \ion{H}{2} regions \citep{bau98}). 
For the models described below which fit the data best for the excitation of the nebula, most of the Fe is in \ion{Fe}{3} and \ion{Fe}{4}. In some cases, the \ion{Fe}{4} fraction is $\sim$ 90 $\%$, but this still requires the bulk of the iron to be condensed on grains if the total abundance is near or above the solar value.

\subsubsection{Recombination Lines and Excitation}

The map of the observed ratio of \he1 21127 \AA \ to Br$\gamma$ is shown in the left panel of Figure~\ref{ratio}. The ratio is approximately uniform across the main emission in the center of the image and has a value of between 0.04 and 0.05. The highest value is 0.055 near the continuum peak, and regions around the periphery have values in the range 0.025--0.03. The value in the average spectrum of Figure~\ref{spec} is 0.046 (Table~2) which includes the effect of the blend of lines near Br$\gamma$. These other lines are not included in the model prediction shown in Figure~\ref{he1brg}, although their effect is small. The average ratio of 0.046 is consistent with an O--star which has \teff \apge 40000 K \citep{han02}. A small correction for extinction gives an intrinsic ratio of 0.051 if we adopt the observed ratio in Table~2. 

Cloudy models were run for a grid of ionizing flux (phi, s$^{-1}$ cm$^2$) versus exciting source \teff. The former parameter was varied from 12.0 to 16.0 dex and the latter varied from 25000 K to 45000 K. 
The results for our baseline case (density $=$ 10$^{4}$ cm$^{-3}$ and blister geometry) are given in Figure~\ref{he1brg}. The ratio along a given contour becomes approximately constant above \teff \ $=$ 40000 K, but the detailed ratio depends on the ionizing flux as well. In Figure~\ref{he1brg}, the overall emission due to dust compared to the Br$\gamma$ flux as predicted by Cloudy is also plotted (as {\it dashed} contours). The intersection of this contour with the \he1 to Br$\gamma$ contour can give an idea of the true excitation of the nebula under the assumption of a single dominant source (which might not be the case); see \citet{bd99} and \citet{bm08}.

The dust emission was calculated approximately from the spectral energy distribution (SED) shown in Figure~2 of \citet{how97}. No attempt was made to correct for the different beam sizes associated with the many distinct data sets which make up the \citet{how97} figure. The most important is the $\sim$ 100 \mic \ point originally reported by \citet{th79} since this is the peak in the SED. 
The \citet{th79} beam was 50$''$ in diameter and included flux from the nearby K3--50B \ion{H}{2} region. 
On the other hand, higher angular resolution data at 450 \mic \  \citep{t06} show a very compact source (unresolved at the 8$''$ beam size) with a similar flux to the SED presented by \citet{th79}.  
\citet{how96} report a Br$\gamma$ flux of 3.2$\times$10$^{-11}$ ergs cm$^{-2}$ s$^{-1}$ corrected for $A_K$ $=$ 2.17 mag. We find a integrated dust emission of approximately 1.8$\times$10$^{-6}$ ergs cm$^{-2}$ s$^{-1}$ from the \citet{how97} SED. Thus the lower limit to the ratio is 1.78$\times$10$^{-5}$.

The intersection of the contours for \he1 and dust compared to Br$\gamma$ suggest a \teff \ of about 38000 K and Log10(phi) $=$ 13.8. Since the Br$\gamma$ to dust ratio is a lower limit according the discussion above, the true \teff \ would be hotter. The ionization model grid was repeated for a density of 10$^5$ cm$^{-3}$. The results are similar to those in Figure~\ref{he1brg} with the intersection of the contours for the observed ratios moving to slightly lower \teff \ and higher ionizing flux, i.e up and to the left in the figure. The formal intersection is at 37000 K and Log10(phi)=14.5. Changing the geometry from plane parallel (open) to spherical (closed) in Cloudy requires higher incident ionizing flux due to the radiation field dilution by $1/r^2$. In the model grid, the inner radius of the cloud was set to 0.5$''$ or 5$\times$10$^{16}$ cm. The observed ratios for the spherical case intersect at somewhat higher \teff, 40000 K, and the corresponding ionizing flux is Log10(phi) $=$ 14.7 to 14.4. If the dust emission is too high by a factor of two due to beam size mismatches, then the Br$\gamma$ to dust ratio is too low by a factor of two. In this case, the model \teff \ could be as high as 45000 K for a spherical geometry. 

We conclude that the excitation of \k350 \ is  consistent with a central star whose \teff \ is greater than 37000 K and less than 45000 K. This is higher than \citet{oka03} derive based on line ratios of collisionally excited ions. A similar result is obtained in W51 IRS2 when comparing the results of \citet{oka01} for line ratios to the recent spectral type obtained for the star located at the position of IRS2 (whose \uchii \ region is known as W51d); see \citet{bar08}. Evidence is presented below that the geometry may be more open \ like, see also][]{col91, depree94} which would lead to \teff \ values near 40000 K. 

\citet{depree94} report a Lyman continuum luminosity, LyC $=$ 2.1$\times$10$^{49}$ s$^{-1}$ for a distance of 8.7 kpc. Adopting 7 kpc (see \S1), results in 
a reduced number of 1.3$\times$10$^{49}$ s$^{-1}$. \citet{kcw94} find LyC $=$ 1.95$\times$10$^{49}$ s$^{-1}$ for a distance of 8.3 kpc, which scaled to 7 kpc gives a similar number. This LyC luminosity is consistent with a single O--type star with effective temperature of approximately 40000 K based on the properties for O stars given by \citet{mar05} or 41000 K if the data from \citet{vac96} are used. Figure~\ref{c217} shows there are multiple continuum sources (especially in the vicinity of the peak emission), but the excitation of the nebula based on the NIFS spectrum is consistent with a single, dominant star.

\subsubsection{\he1 20587 \AA \ Emission} 

The strongest \he1 line in \k350 \ is the 20587 \AA \ line (Figure~\ref{spec}).  The right panel of Figure~\ref{ratio} shows the line ratio relative to Br$\gamma$. Comparison to the left panel of Figure~\ref{ratio} indicates the 20587 \AA \ line emission is more variable across the nebula than is the case for the 21127 \AA \ emission. The strong degree of uniformity for the latter suggests the 20587 \AA \ line variations are not due to differential extinction, but instead are due to structure in the nebula. If the variations seen in the right panel of Figure~\ref{ratio} were due only to variations in the line of sight extinction, then this would require a change in A$_K$ of more than 10 magnitudes. The continuum map morphology is not consistent with this picture, nor is the average extinction found by \citet{how96} in the central few arcseconds (A$_K$ $\sim$ 2). Finally, the Br$\gamma$ and \he1 21127 \AA \ intensities are near maxima at the position of the continuum peak while the 20587 \AA \ line itself is weak there. 

The formation of the \he1 20587 \AA \ line is complicated and depends on the physical conditions in the ionized gas \citep{s93, fer99}\footnote{The numerical values of the 20587 \AA \ line have changed since the \citet{s93} paper; see \citet{bd99, fer99}}. A large ratio to Br$\gamma$ is due primarily to high density \citep{s93} which allows for enhanced collisional population of the 2 $^{1}$P$^{\circ}$ level. However, this level population can also depend strongly on \ion{He}{1} L$\alpha$ 584 \AA \ resonance fluorescence which is affected by dust absorption of the He L$\alpha$ line and ionization of neutral H by the He L$\alpha$ line since both processes destroy the photons which would otherwise pump the line \citep{s93}. The biggest changes in 20587 \AA \ line formation are expected for a range of central star \teff. The \citet{s93} models were for an average over the nebula; in the present case, the nebula is spatially resolved and clearly is dominated by a point source (or unresolved sources) at one end of the FOV. 
In Figure~\ref{depth}, the ratio of \he1 to Br$\gamma$ emissivity with depth into the cloud (for a blister model) is plotted for the 21127 and 20587 \AA \ lines and a central source with \teff \  $=$ 38000 K. The 20587 \AA \ line strength changes sharply from inner (close to central source) to outer (far from the central source) positions in the cloud. 

Thus, at any given point in the nebula, we would expect a relatively high ratio in the right panel of Figure~\ref{ratio} since the effects shown in Figure~\ref{depth} would be integrated along the line of sight. The range of observed values for the ratio in the right panel of Figure~\ref{ratio} then suggests that real density variations may exist across the field of view (lower density near the continuum peak). A possible exception to this would be if we were looking along a line of sight such that a slab or blister was illuminated with a geometry whose normal to the slab was in the plane of the sky. But this geometry is inconsistent with  the molecular hydrogen morphology (see below, \S5.4) which suggests a face--on blister geometry. 

In Figure~\ref{density}, the cloudy predictions for the ratios of the 20587 \AA \ and 21127 \AA \ lines to Br$\gamma$ versus density  from Cloudy are plotted. The latter line has an observed ratio of approximately 0.051 (corrected for extinction) and is nearly uniform over the nebula. The former line has an observed ratio (corrected for extinction) that varies from about 0.33 to 0.84. Figure~\ref{density} shows the Cloudy prediction for the case of \teff \ $=$ 38000 K and Log10(phi) $=$ 14.0. These results indicate that the 21127 \AA \ line ratio is nearly constant over a range in density of 100 to 100000 cm$^{-3}$ while the 20587 \AA \ line changes from 0.38 to nearly 1.0 over the same range. The ratios plotted in Figure~\ref{density} are integrated along the line of sight through the nebula in contrast to the ratios plotted in Figure~\ref{depth}.

\subsection{Kinematics}

The kinematics evident in the bright Br$\gamma$ line (Figure~\ref{velmap}) do not fit neatly with any of the previous kinematic data that were obtained at lower angular resolution. Figure~\ref{velmap} shows a double lobed structure, possibly a flow, with a major axis distinct from any of the larger scale components discussed above. Furthermore, the ordered red and blue shifted material is not well correlated with much of the strong Br$\gamma$ emission and the point of symmetry of the indicated flow is not coincident with the continuum peak. A possible explanation for the small scale flow is that it is created by a lower mass protostar in the nascent cluster. This molecular outflow is then ionized by the massive star at the continuum peak.  A similar scenario was recently described for the compact high mass star forming region IRS2 in W51 \citep{lacy07, bar08}. 

Another possibility is that the large scale ionized flow is related to the inner flow. This would require that the outflow is precessing or that details of the interaction of the out-flowing gas on larger scales than the NIFS FOV reorient the flow since the axis of the inner flow does not line up with the axis of the large scale ionized flow.  The contours of the larger flow do, however, appear to twist \citep{depree94} and Figure~\ref{c217} clearly shows more detailed radio (kinematic) observations would be helpful in understanding the transition between the inner scale and larger scale gas kinematics.  \citet{depree94} argued that the large scale ionized flow might result from the expansion of material coming from within a small (few arcsecond) shell. They took as evidence for a shell the high angular resolution 2 cm map of \citet{tm84}. The NIFS Br$\gamma$ morphology appears to match well with the 2 cm emission, but does not appear shell like \citep[][noted the classification of \k350 as shell like was problematic]{tm84}. Still, the NIFS data suggest an outflow originating within this small region which is likely confined or directed by the material around it. 
The large scale flow has a velocity gradient of about 6 \kms \ arcsec$^{-1}$ while the small scale flow gradient evident in Figure~\ref{velmap} is \apge 25 \kms \ arcsec$^{-1}$. The velocity reported for the core of the radio emission is $-$33 \kms \  with respect to the LSR \citep{depree94}. The integrated Br$\gamma$ emission has $V_{\rm LSR}$ $=$ $-$62.5 \kms \ (see Table~1). The red--shifted emission in Figure~\ref{velmap} is at $V_{\rm LSR}$ $\sim$ $-$30 \kms. 
This suggests that we may be only seeing part of radio--emitting gas in the Br$\gamma$ map and that the red lobe of the Br$\gamma$ flow is really tracing the bulk of the radio flow; that velocity is presumably the systemic velocity of the radio \uchii \  region. In this model, all the rest of the Br$\gamma$ emission is tracing blue--shifted gas outflowing from the central source. The spatially unresolved Br$\gamma$ emission at the continuum peak (see Figure~\ref{brgmap} and \S 5.6) also has a $V_{\rm LSR}$ $\sim$ $-$32 \kms \ that is close to the systemic velocity defined by the H76$\alpha$ emission.  

\subsection{Nebular Structure}

In \S 4.3, a broad component of emission was identified to the blue of the Br$\gamma$ line (see Figure~\ref{detail}). The position of the line is consistent with one of the \he1 lines from the n$=$ 7--4 complex, but the emission is much stronger than expected based on transition probabilities and broader than any other \he1 line by about a factor of two. Thus, it is more likely a second Br$\gamma$ component. The line velocity is about 40 \kms blue shifted compared to the background or cloud velocity (the green regions in Figure~\ref{velmap}). This is larger than the blueshifted velocity structure of the principal component of Br$\gamma$ shown in that Figure. The width of the broad component is about twice the width of the principal Br$\gamma$ component (120 \kms \ compared to 70 \kms). The broad component is distributed in an arc to the SSE of the continuum source ($\sim$ 0.5$''$ in length and offset from the peak by $\sim$ 0.2$''$--0.3$''$) and is confined to this vicinity of the continuum source. 

It is not clear how the high velocity, broad component is produced. It may be related to hot gas escaping a tightly constrained volume near the central source. The scale associated with the broad component is small, only 0.007--0.010 pc (for a distance of 7 kpc). In champagne flows that are produced as the ionizing radiation from the hot star penetrates into a uniform medium \citep[e.g.,][]{york84}, the hot star quickly carves out a larger diameter cavity (\apge 0.1 pc). Perhaps shocked gas initially flowing parallel to the plane of the sky and then being redirected along our line of sight could explain the broad component. This putative flow might be related to the central source stellar wind. 

In any case, gas must escape to the north and eventually form the larger ionized flow seen in the \citet{depree94} radio images. The immediate vicinity of the continuum peak may be filled with low density gas, and this gas may surround (or overlay) higher density material within the projected "cone" which extends to the south of the continuum source. In order to explain the low ratio of \he1 20587 \AA \ to Br$\gamma$ near the continuum peak, the particular line of sight must contain very little dense ionized gas indicating this area has been cleared away as the hot star emerges from its parent cloud (which remains mostly to the north). The high angular resolution 2 cm image of \citet{tm84} and the molecular maps of \citet{how97} are consistent with clearing of high density material in the vicinity of the continuum peak; both show voids or gaps in the emission.  None of the current absolute positions is good enough to place the different wavelength data to better than about 0.5$''$ which is the scale of the structure in our data and the 2 cm map. 

The nebula is seen in projection against H$_2$ emission, primarily in the lower left quadrant where the Br$\gamma$ emission is strongest (Figure~\ref{h2}) but also (unresolved) at the position of the continuum peak. Three prominent lines of molecular hydrogen are listed in Table~1. The average ratios of the two longer wavelength lines to the v$=$1--0 S(1)  21218 \AA \ line (0.42 $\pm$ 0.1, 0.23 $\pm$ 0.1 for H$_2$ v$=$ 1--0 S(0) and H$_2$ v$=$ 2--1 S(1), respectively) can be compared to the diagnostic diagram given by \citet{han02} which shows these same ratios for a large sample of \uchii \ regions along with several model predictions for shock excited and fluorescent emission. The bulk of the \citet{han02} sample is consistent with dense (10$^6$ cm$^{-3}$) photo--dissociation regions (PDR). The results for \k350 \ place it in the same region of the diagram showing it too is consistent with a dense PDR on the boundary of its ionized nebula. Since we see both the ionized gas and molecular emission, it is likely the H$_2$ emission is behind the ionized gas (a large column of molecular gas in front would obscure our view). If this is the case, then the \uchii \ region may be best described as blister like; the hot star has emerged from the parent cloud and is ionizing the face of the molecular material to the south.  In this case, the "cone" is really a flattened structure seen face on and lying on the surface of a dense cloud of molecular gas.

\subsection{The Ne, Ar, and S Lines}

As mentioned above, \citet{oka03} analyzed [\ion{Ar}{3}]/[\ion{Ne}{2}] and [\ion{S}{4}]/[\ion{Ne}{2}] ratios in \k350. The observed ratios are consistent with other \uchii \ regions, but inconsistent with simple ionization models. The models predict stronger [\ion{S}{4}]/[\ion{Ne}{2}] than [\ion{Ar}{3}]/[\ion{Ne}{2}] for a given \teff, but the observations show the average ratios are of similar value and the peak ratios have substantially larger [\ion{Ar}{3}]/[\ion{Ne}{2}] than [\ion{S}{4}]/[\ion{Ne}{2}]. Our grid of models agrees with \citet{oka03} in the sense that for a given \teff, the [\ion{S}{4}] line ratio is predicted to be stronger. 

\citet{mart02} computed Ar, S, and Ne abundances for \k350 \ from ISO mid--infrared spectra and find abundances which differ from the solar values (and our Cloudy models). In particular, the Ne abundance compared to H is approximately solar, while the Ar abundance is a factor of 2 lower than the solar value, and the S abundance is a factor of 10 lower. However, this large reduction in S abundance may be (in part) due to the high density in \k350 \ depopulating the upper levels producing the S emission as \citet{mart02} point out. Detailed models used to derive the S abundance in \k350 \ by \citet{aff97} give a value only a factor of two lower than solar. The ISO spectra were obtained through large apertures ($\sim$ 20$''$), and the line ratios are smaller than the peak ratios in the higher angular resolution \citet{oka03} data. The ISO line ratios are nearly equal and have a value of approximately 0.25. This is comparable to the low end of the range of values obtained by \citet{oka03}. The observed high peak line ratios of \citet{oka03} ([\ion{S}{4}]/[\ion{Ne}{2}] $=$ 3, [\ion{Ar}{3}]/[\ion{Ne}{2}] $=$ 4) would appear to require lower S abundance and higher \teff \ to be consistent with the Cloudy ionization models, but the precise line ratios do not converge in our grid of models even when adopting the detailed abundances of \citet{mart02}. In particular, reproducing the high line ratio of [\ion{Ar}{3}]/[\ion{Ne}{2}] requires \teff \ to be $\>$ 45000 K, the limit of the range based on the analysis of the \he1 21127 \AA \ line and the dust emission, and such a model would produce a much too high [\ion{S}{4}]/[\ion{Ne}{2}] ratio. 

Earlier work based on Kuiper Airborne Observatory observations \citep{col91} also did not succeed in fitting the ratios of these infrared lines. The \citet{col91} data were obtained through apertures approximately 5$''$ in diameter and the line ratios are larger than the ISO values (about unity for both) and somewhat smaller than the \citet{oka03} peak values. \citet{col91} and \citet{oka03} have argued that multiple ionizing sources might explain the observations. 

The NIFS data generally (continuum, \he1 lines, resolved nebular structure) seem to indicate the central few arcseconds are dominated by the continuum peak. Figure~\ref{feiii} shows that the excitation in the nebula is higher toward the continuum peak. The [\feiii] emission is weak in the vicinity of the continuum peak compared to the [\seiv] line. The [\kriii] line is weak as well and this is consistent with its lower ionization potential than the other species present.  In the case of OKYM4 (Figure~\ref{c217}), there may be some evidence for lower excitation; \citet{oka03} preferred a late O or B star for this source. The left panel of Figure~\ref{ratio} shows a dip in the \ion{He}{1} 21127 \AA \ to Br$\gamma$ ratio at the position of OKYM4 (the ratio is about 0.025). This source is about 2$''$ away from the continuum peak and this is roughly the scale of an associated  Stromgren sphere for the central star (Figure~\ref{depth}). Further, the [\feiii] emission peaks at the location of this source. Indeed, the emission morphology of the line maps suggest the boundary of the \ion{H}{2} region which is principally due to the continuum peak is on this same scale. If this is correct the ``cone'' morphology arises from dust obscuration to the north and the ionization boundary of the nebula to the south. The main nebula is largely excited by the continuum peak, and OKYM4 may be locally internally excited. The compactness of OKYM4 then suggests the source is very young and/or highly constrained.

Lower mass OB stars should not greatly affect the structure within the ``cone,'' however. \citet{col91} and \citet{oka03} argued that the low {\it average}  [\ion{Ar}{3}]/[\ion{Ne}{2}]  and [\ion{S}{4}]/[\ion{Ne}{2}] ratios might be due to multiple lower \teff \ sources (as may be the case for OKYM4), but a successful model must also explain the detailed ratios including their very high peak ratios near the continuum peak. While we have not been able to produce such a model, higher ratios are more consistent with hotter stars.

\subsection{The Continuum Peak}

In the preceding sections, the structure of the nebula was described, and it was concluded that the central source is a hot star emerging from its natal cocoon. The star is ionizing the cloud to the south in our FOV and produces an intense \uchii \ region and associated PDR. It was argued that the density in the nebula may be lower in the vicinity of the continuum peak. 

The continuum peak (by definition) dominates the $K-$band emission. In addition, Br$\gamma$, \he1 \ 21127 \AA, [\ion{Se}{4}] (the highest ionization line observed), and H$_2$ show unresolved emission peaks (or near peaks) at this location. There is clearly a very compact source of emission here which includes stellar continuum (from multiple sources) re-radiated by hot dust, and possibly line emission from the circumstellar environment. An ionized circumstellar disc is a possible source of the compact emission. Though no kinematic evidence is available to support this possibility. \citet{bik04} and \citet{blum04} presented kinematic evidence of discs around (somewhat less) massive young stars based on the emission profiles of the CO v$=$2--0 rotational--vibrational bandhead at 22935 \AA. This line is not detected in \k350. Whatever its geometry, there must be compact circumstellar material (partially) surrounding the central source which is not directly linked (at least now) to the larger resolved \uchii \ region covered by the NIFS FOV.

\section{Summary}

High angular resolution ($\sim$ 0.2$''$) $K-$band spectral imaging has been presented for the Galactic \uchii \ region \k350. The data were obtained with adaptive optics and the image slicing integral--field spectrograph, NIFS, at Gemini North. The combination of excellent wavelength coverage, moderate spectral resolution (5160), and high angular resolution provide the most detailed look to date at this massive young stellar object and its immediate environment. The emission morphology matches well with earlier speckle observations.

Spatial and ionization structure are resolved in the 3$''$ nebula. The \he1 emission morphology suggests changes in density across the field which are attributed to the break out and clearing phase of massive star birth. The ratio of \he1 20587 \AA \ to Br$\gamma$ varies from about 0.3 to 0.8 and requires that the density near the central source be more than a factor 10 less than farther away. The ionization in the nebula can be traced across the nebula by the varying intensity of [\ion{Fe}{3}] and [\ion{Se}{4}] lines. 

A grid of ionization models has been run with the code Cloudy \citep{fer98}, and the results of these calculations were compared to the observed \he1 21127 \AA \ line to Br$\gamma$ line ratio and dust emission properties. The combination of observed values constrain the nebular excitation to a source with \teff \  \apge \ 37000 K. This is somewhat hotter than expected from collisionally excited emission--line ratios in the mid--infrared \citep{oka03}, but those line ratios, particularly their peak values on scales of $\sim$ 1$''$, are not explained by our models \citep[or those of][]{oka03}. 

Despite the high angular resolution of these data which enhances the contrast between background and point source emission, no photospheric lines are detected in the spectrum of  the bright continuum source; it remains buried within intense continuum emission. The central source is likely also blended with other point sources as pointed out by other investigators. There is spatially unresolved, bright emission in the lines of Br$\gamma$, \he1, and [\seiv] (and possibly [\feiii]) at the continuum peak position and these are attributed to the circumstellar environment of the massive young stellar object.  

The NIFS data cube reveals a striking kinematic signature in \k350. However, this ``flow'' is not aligned with the large scale ionized out flow reported in earlier radio continuum and recombination line studies (10's of arcseconds in extent).  The NIFS kinematic signature is not symmetric about any detected point source in the NIFS image either. It is possible the bipolar kinematics arise from a lower mass protostar associated with the central star of \k350,  and the associated material is ionized by the hot central source. A second broad component of Br$\gamma$ emission is seen within a few tenths of an arcsecond of the continuum peak. This may be hot gas escaping from the cavity resulting from the action of the ionizing radiation on the natal molecular cloud or large scale torus.

\begin{figure}
\includegraphics[width=6.7 in]{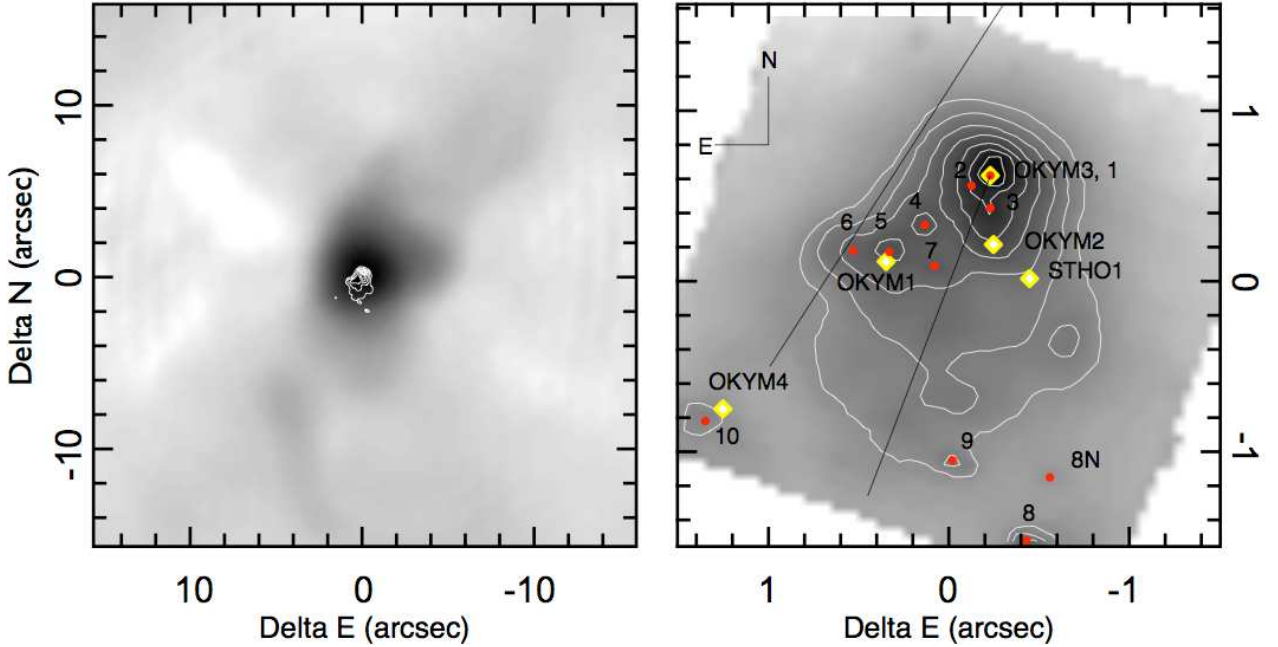}
\caption{ 
Continuum maps. {\it Left} panel: Log of the intensity grayscale for the 14.7 GHz radio continuum image from \citet[see their Figure 1]{depree94}.  The angular resolution is 2.0$''\times1.6''$ (beam PA$=$79$^{\circ}$). The contours correspond to the NIFS continuum ({\it right} panel) where the peak emission in both images is made to match. {\it Right} panel:
$K-$band continuum at 21700 \AA \ (19.2 \AA, 9 pixels wide) shown in grayscale (log intensity) and with 
the same contours plotted in the {\it left} panel. The field center is located at approximately
RA(2000)$=$20$h$~01$m$~45.71$s$, Dec(2000)$=+$33$d$~32$'$~42.1$''$
(see text for a discussion of source position versus wavelength). The
position angle on the sky was 70\deg (east of North); this image has been rotated so that North is up, and East is left. The {\it red} filled circles are sources from \citet{hof04} and the {\it yellow} diamonds
are the mid--infrared sources of \citet{oka03}; see text. The continuum peak appears as an unresolved point source near 0.2, 0.6 and is assumed to be Source \#1 of \citet{hof04} and OKYM3 of \citet{oka03}.
The long line starting at Source \#1 indicates the approximate direction of  the ionizied outflow described by \citet{depree94}. The second line is our estimate of the rotation axis of the
HCO$^+$ torus described by  \citet{how97}; see text. The torus is centered a few arcseconds N-W of the continuum peak.
\label{c217}}
\end{figure}

\begin{figure}
\epsscale{1.}
\plotone{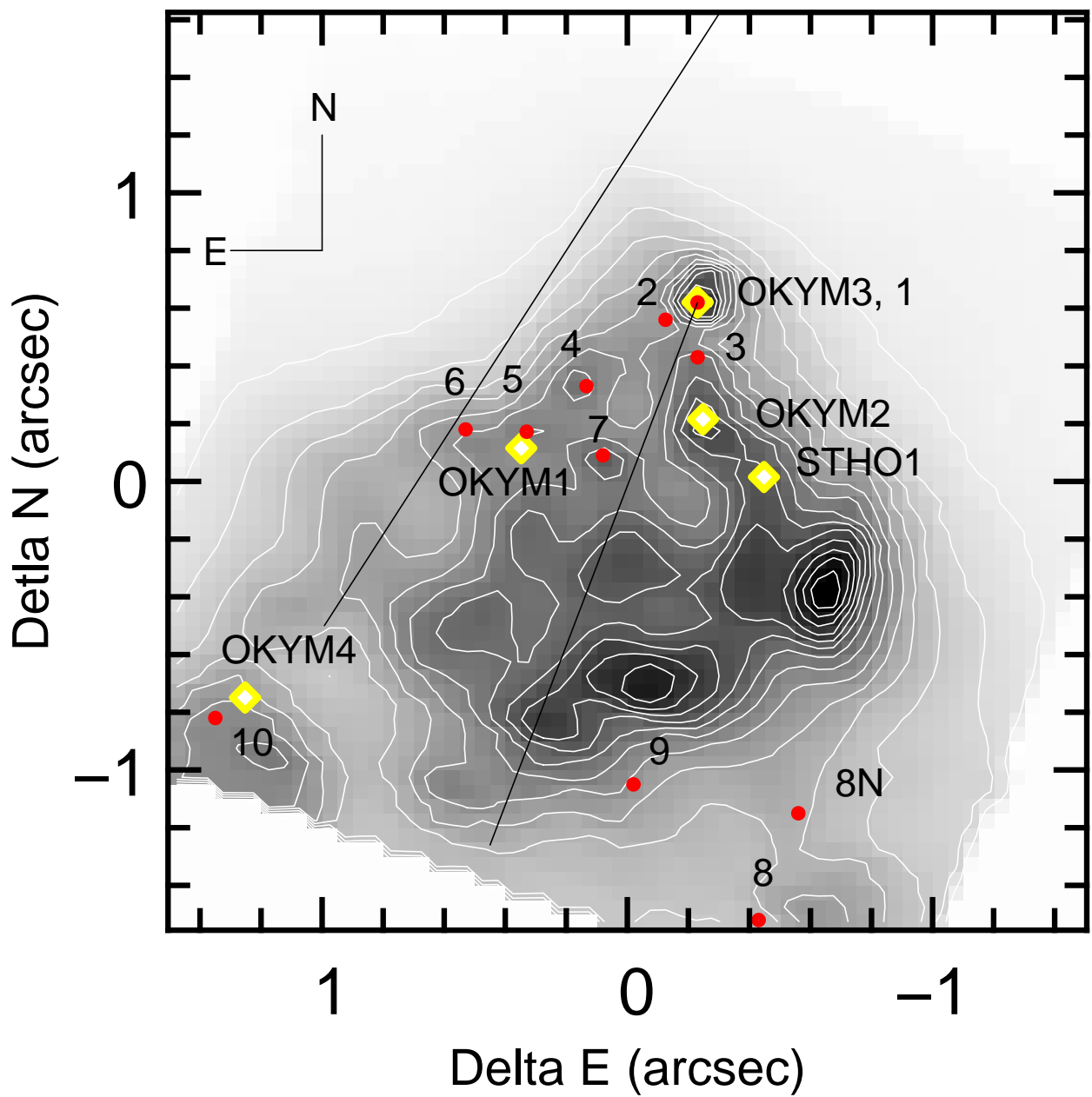} 
\caption{The continuum subtracted Br$\gamma$ map. The lines and
  points are the same as for Figure~\ref{c217}. Several of the continuum
  sources detected in Figure~\ref{c217} are among the brightest Br$\gamma$
  sources. The brightest peaks appear angularly resolved, and it is not
  clear if there are associated point sources embedded within or if
  these are ionized clumps of higher density material. There are 14
  contours overlaid on the image. The lowest contour is
  1.48$\times$10$^{-13}$ erg cm$^{-2}$ s$^{-1}$ arcsec$^{-2}$ and the contour
  spacing is the same. The peak is located at  $-0.6,-0.4$
  and has a value of approximately 2.20$\times$10$^{-12}$ erg cm$^{-2}$
  s$^{-1}$ arcsec$^{-2}$; see text. The Br$\gamma$ line is affected by \he1 \ blends; see text.
\label{brgmap}}
\end{figure}

\begin{figure}
\epsscale{1}
\plotone{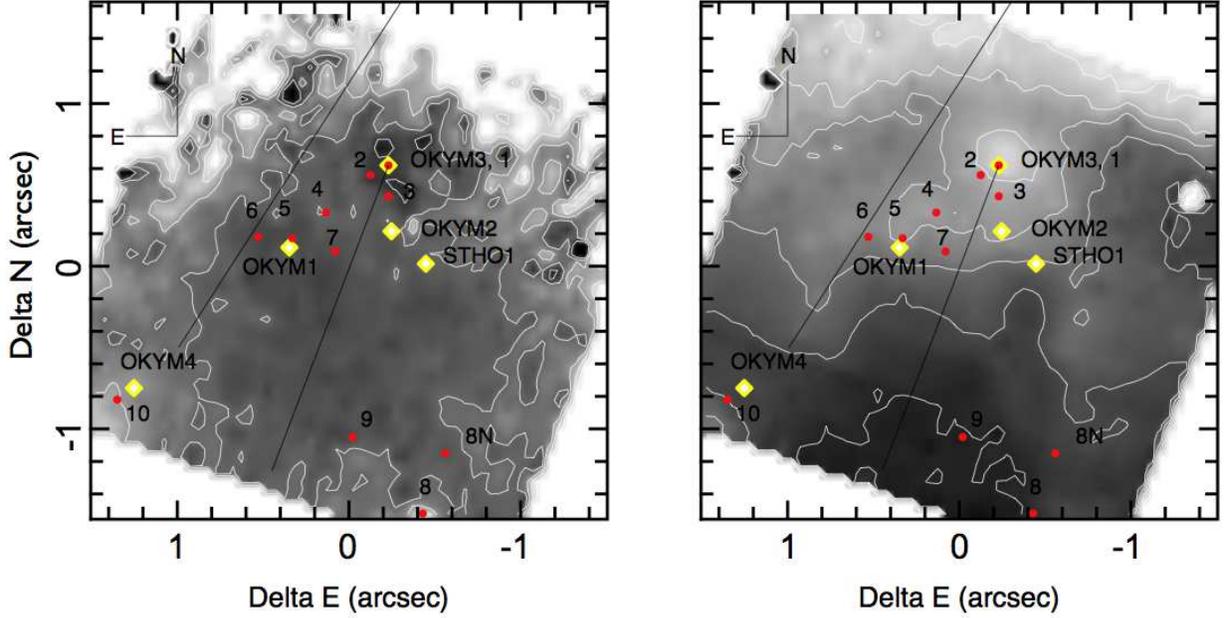} 
\caption{The ratio of \he1  to Br$\gamma$ (grayscale and contours). The lines and
points are the same as for Figure~\ref{c217}. {\it Left} panel, $3p-4s$ $^3$P$^{\circ}$--$^3$S 21127 \AA \ line. The contours are spaced by 0.01 and the peak ratio near
Source \#1 is 0.055. The ratio (uncorrected for extinction) is uniform over most of the emission region and has a value of approximately 0.04 to 0.05.  \he1 blends can affect the strength of Br$\gamma$ as does a broad line component near the continuum peak; see text and also Table~2. The area near OKYM4 has a smaller ratio, $\sim$ 0.025, and this may be evidence of a lower excitation source. It also suggests the dominant continuum peak ionizes a volume corresponding to about 2$''$ (see also Figure~\ref{depth}). {\it Right} panel, \he1 $^1$S--$^1$P$^{\circ}$ 20587 \AA \ line. In contrast to the 21127 \AA \ line, this ratio shows strong variations over the FOV: lower values ($\sim$ 0.25) at the continuum peak location and systematically increasing away from the peak to the south ($\sim$ 0.7) in the panel. \label{ratio}}
\end{figure}

\begin{figure}
\includegraphics[width=6.5 in]{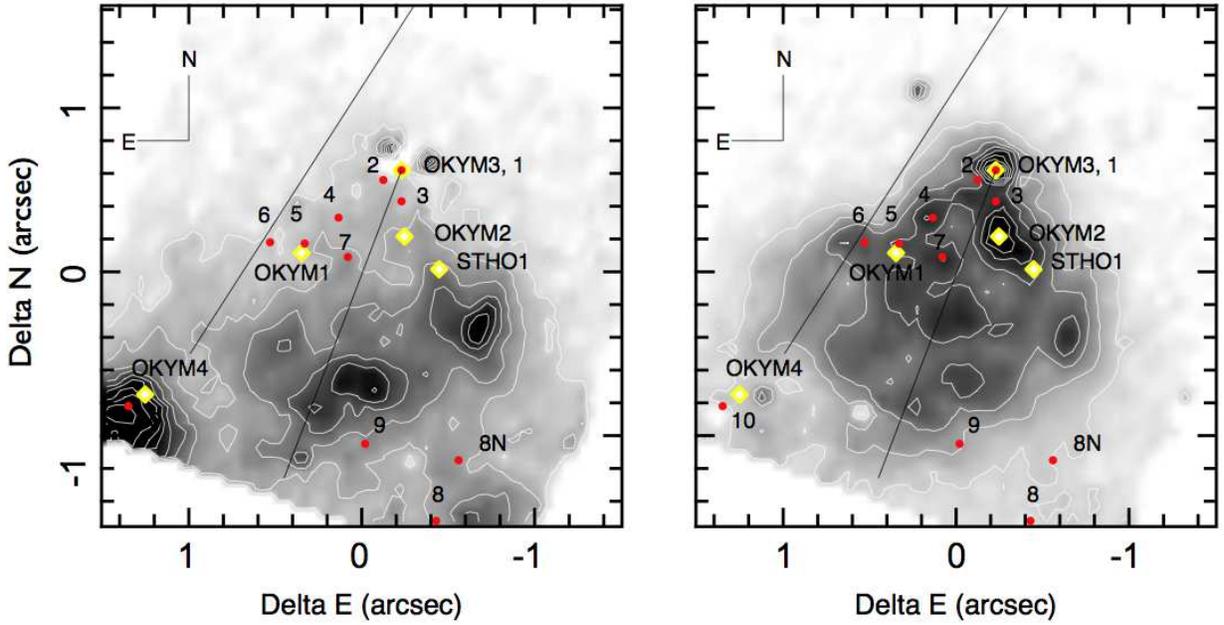}
\caption{Higher ionization line maps. The lines and
  points are the same as for Figure~\ref{c217}. [\feiii] 22184 \AA \ is shown on the left, and [\seiv] 22867 \AA \ is shown on the right. The ionization potential of \feiii \ is lower than that for \seiv \ showing that the excitation is higher closer to the continuum peak. The peak surface brightnesses are 6.6$\times$10$^{-14}$ erg cm$^{-2}$ s$^{-1}$ arcsec$^{-2}$ and 1.1$\times$10$^{-13}$ erg cm$^{-2}$
  s$^{-1}$ arcsec$^{-2}$  for [\feiii] and [\seiv], respectively, and the contour intervals are evenly spaced at an interval 10 times less than the peak value.
\label{feiii}}
\end{figure}

\begin{figure}
\plotone{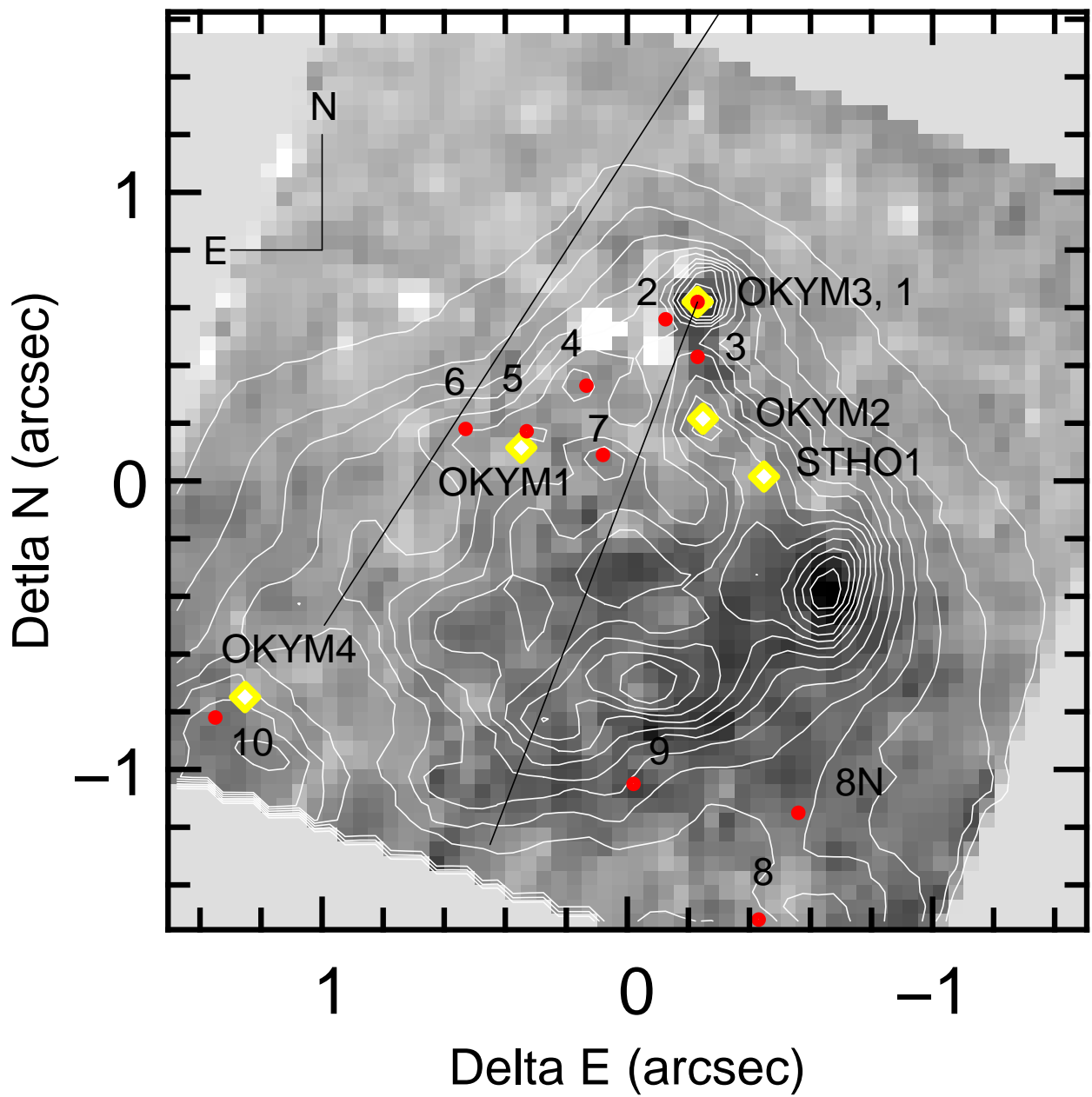}
\caption{Molecular hydrogen emission map in the v$=$ 1--0 S(1) 21218 \AA \ line. The lines and
  points are the same as for Figure~\ref{c217}. The emission is clumpy and appears to trace the ionized gas (Figure~\ref{brgmap}). The peak emission is centered on the peak of the Br$\gamma$ emission (shown as contours; see Figure~\ref{brgmap}). An equally strong peak is coincident with the continuum peak (Figure~\ref{c217}). The peak value of H$_2$ emission is 1.7$\times$10$^{-14}$ erg cm$^{-2}$
  s$^{-1}$ arcsec$^{-2}$\label{h2}}
\end{figure}

\begin{figure}
\plotone{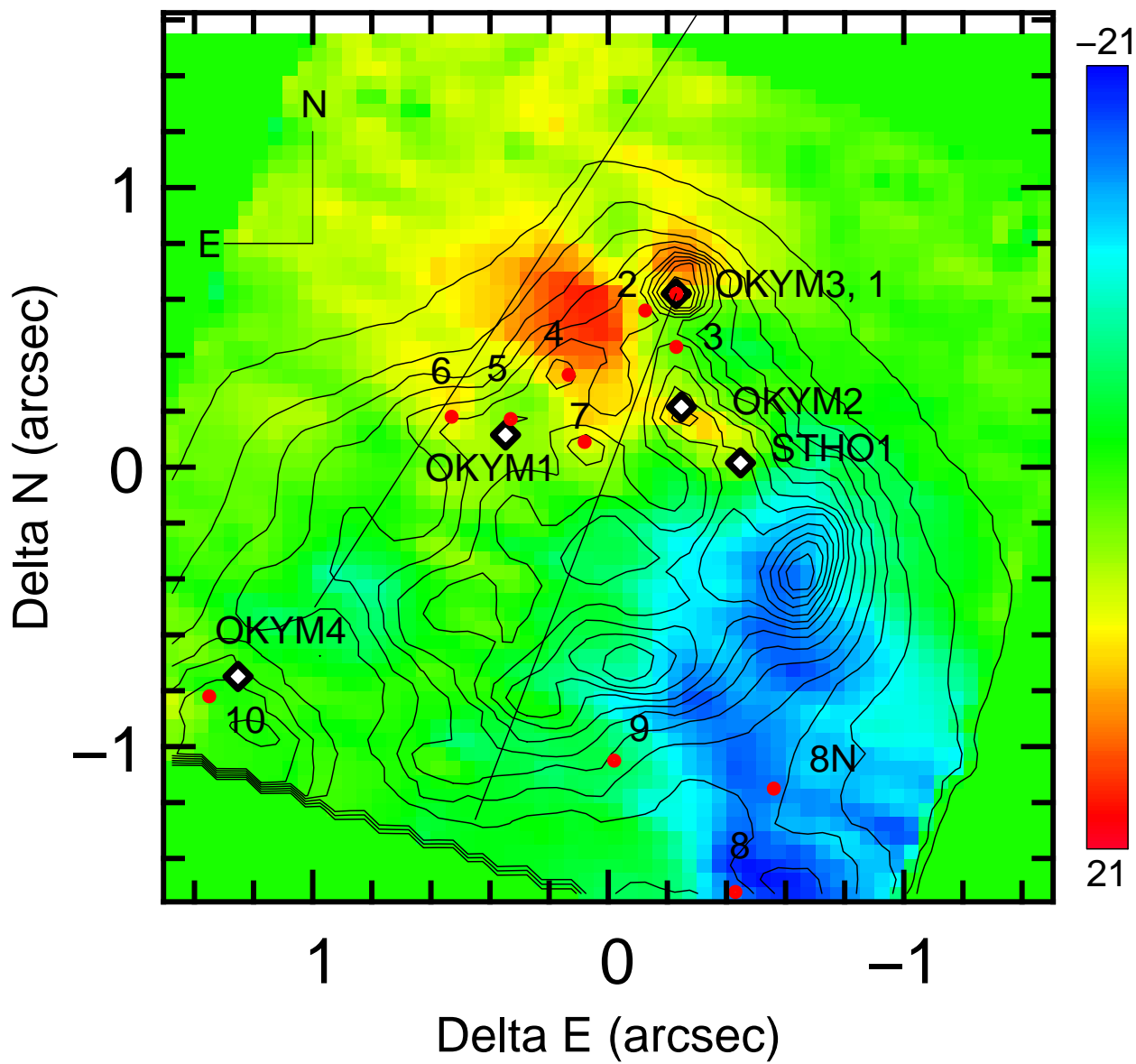}
\caption{Velocity (\kms) for the principal component of the Br$\gamma$ line in each NIFS spaxel derived from the data with a Gaussian fit to the line profile. The labeled lines and points are the same as for Figure~\ref{c217}. 
The mean Br$\gamma$ velocity ($V_{\rm LSR}$ $=$ $-$62.5 \kms) of the principal component has been removed from the map. The bipolar structure does not line up with any known point source, and the axis  does not align with the direction of the large scale ionized flow detected at radio wavelengths. The peak relative velocities are $-$24 \kms \ and $+$31 \kms, but the color map is ``stretched'' slightly to show smaller velocities.
\label{velmap}}
\end{figure}

\begin{figure}
\plotone{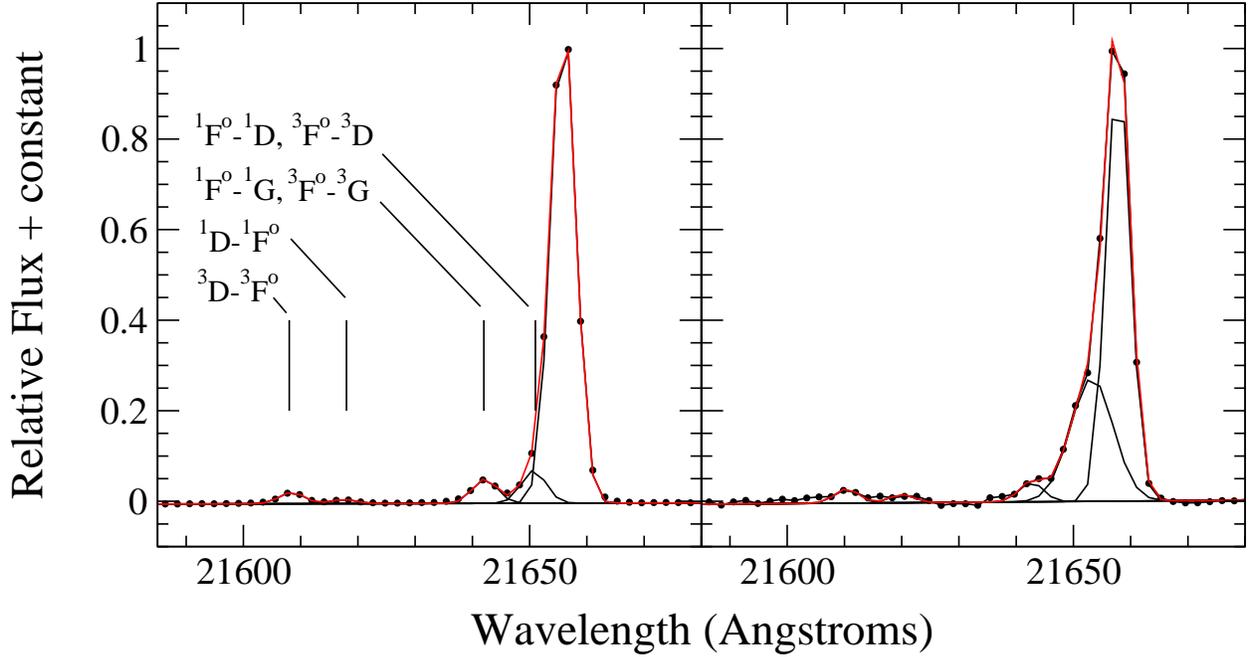}
\caption{Detail of the line emission in the vicinity of Br$\gamma$. The vertical hash marks identify \he1 lines (see Table~2). Five lines were fit simultaneously in each spectrum; see text. {\it Left} panel: emission from a 0.05$''$ aperture located near \citet{hof04} source \#6. {\it Right} panel: emission approximately 0.2$''$ south and east from the continuum peak in an $\sim$ 0.05$''$ aperture. 
The spectra have been continuum subtracted and scaled to the peak Br$\gamma$ flux.
Each panel shows the observed spectrum in {\it filled} circles and the individual lines and continuum as solid {\it black} lines. The combined fit is shown as a solid {\it red} line. The broad line blended with Br$\gamma$ in the right panel may be a distinct blue shifted Br$\gamma$ component. It is much too strong to be due to \he1 under typical conditions; see text. The broad line in the right panel is about 120 \kms  FWHM compared to 70 \kms \ for the principal component of Br$\gamma$ (i.e. the bright component).
\label{detail}}
\end{figure}

\begin{figure}
\plotone{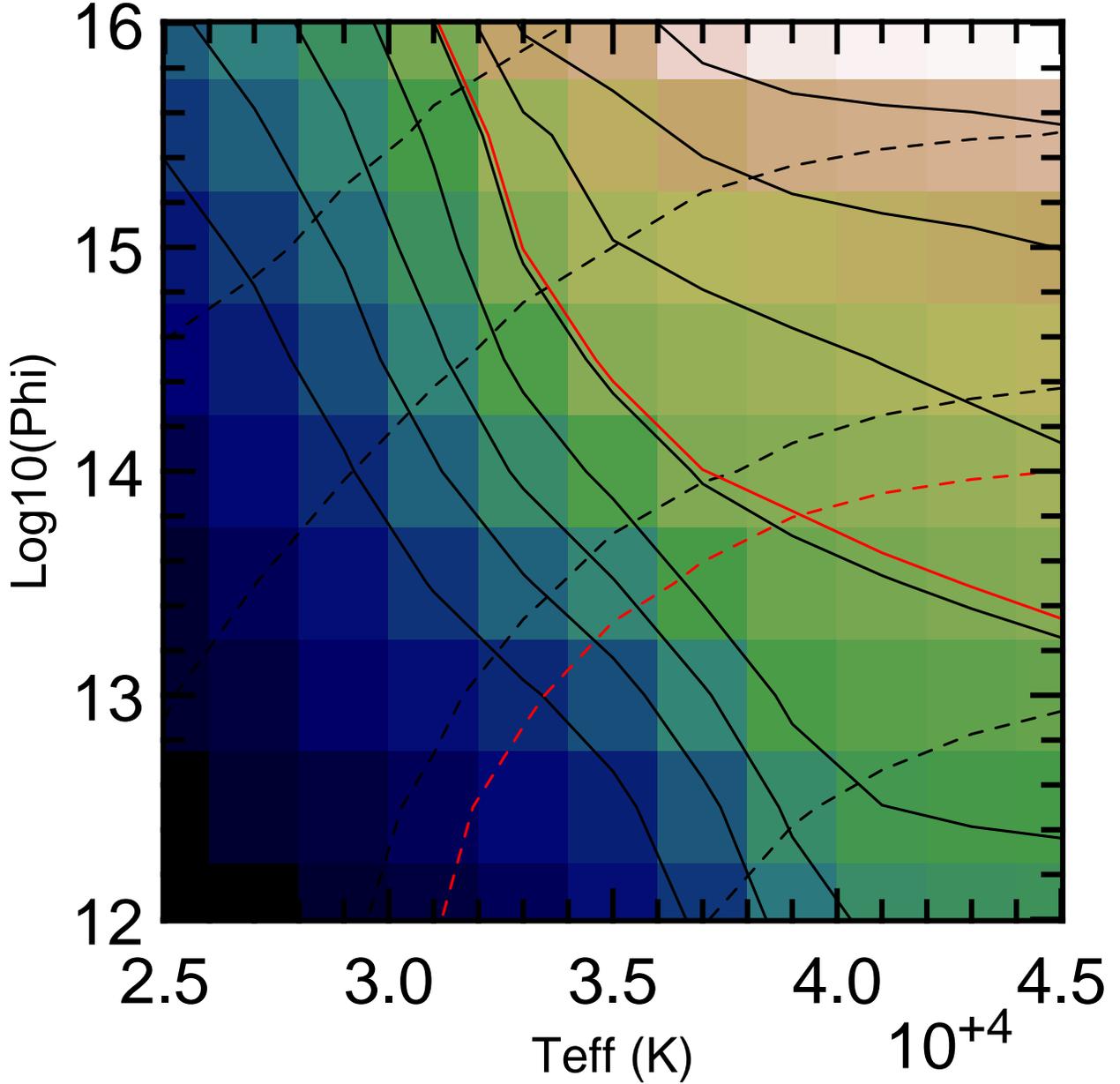}
\caption{Cloudy ionization model prediction of the \he1 21127 \AA \ emission compared to Br$\gamma$ (color--scale and solid contours) for a model grid of \teff \ versus ionizing flux from the central star at the irradiated face of the \ion{H}{2} region. The ratio increases toward higher \teff \ and ionizing flux. The contours have values of 0.01 to 0.08 and are spaced by 0.01. The solid {\it red} contour is the observed value (Table~2) increased by a few percent to 0.049 to account for differential extinction between the \he1 line and Br$\gamma$. The {\it dashed} contours represent the ratio of Br$\gamma$ to total dust emission and decrease toward lower \teff \ and higher ionizing flux. These contours range from 10$^{-7}$ to 10$^{-4}$, each separated by a factor of 10. The dashed {\it red} contour is the observed ratio of 1.8$\times$10$^{-5}$ ($=$ 3.2$\times$10$^{-11}$ erg cm$^{-2}$ s$^{-1}$ / 1.8$\times$10$^{-6}$erg cm$^{-2}$ s$^{-1}$) with Br$\gamma$ corrected for extinction; see text.
\label{he1brg}}
\end{figure}

\begin{figure}
\plotone{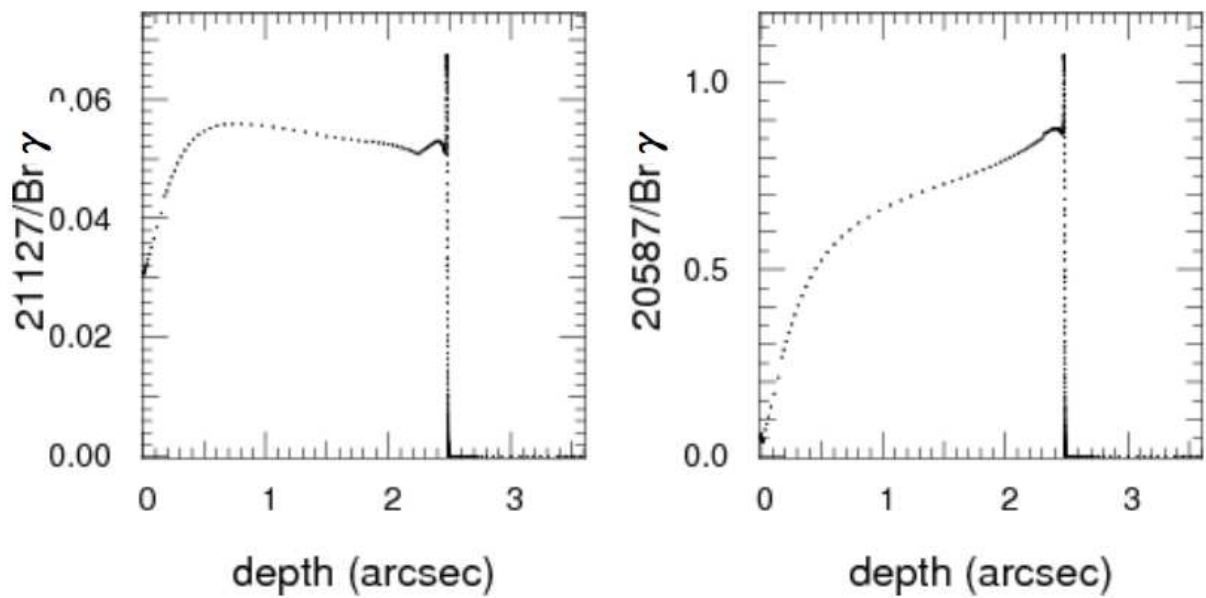}
\caption{The ratio of \he1 to Br$\gamma$ volume emissivities showing where the lines are formed in the cloud. This plot is for a central object with \teff \ $=$ 38000, Log10(phi) $=$ 14, and density 10$^{4}$ cm$^{-3}$. Even though the 20587 \AA \ ratio changes significantly within the cloud, average values of the line ratio integrated through the cloud will be relatively large unless the density is low; see text and Figure~\ref{density}. The 21127 \AA \ and 20587 \AA \ ratios integrated through the cloud are 0.052 and 0.601, respectively. The depth coordinate corresponds to the Cloudy model, converting cm to arcseconds assuming a distance of 7000 pc.
\label{depth}}
\end{figure}

\begin{figure}
\plotone{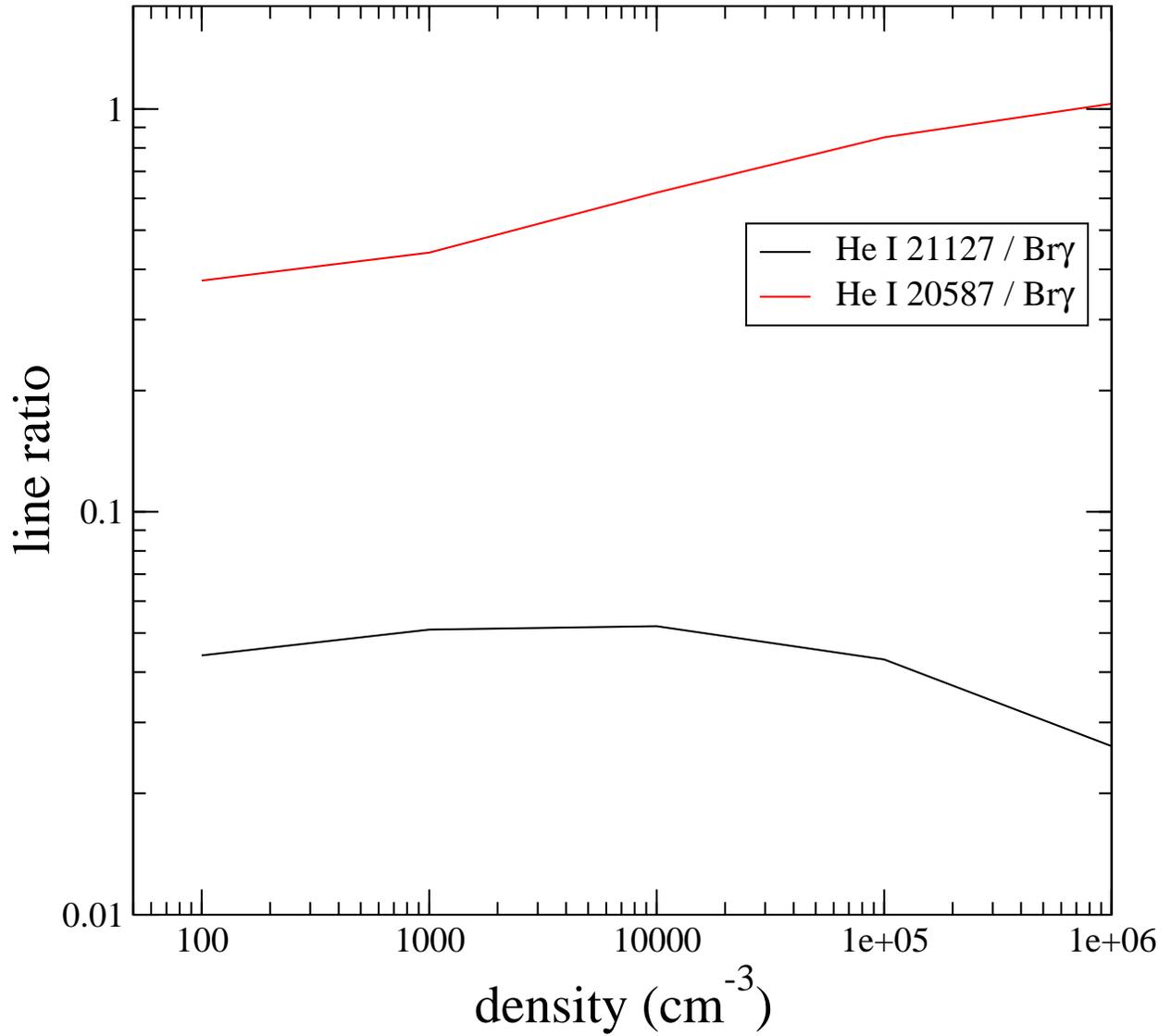}
\caption{The predicted ratio of \he1 lines to Br$\gamma$ from Cloudy for a range of nebular densities. These models correspond to a He abundance by number of 0.1 to H. The models were computed with a central star with \teff \ of 38000 K and the incident radiation field had a flux of ionizing photons, Log10(phi) $=$ 14.0.
\label{density}}
\end{figure}

The authors would like to thank Gary Ferland, Ryan Porter, and Peter van Hoof for useful input regarding Cloudy  and the \he1 model. The authors also thank Manuel Bautista for useful conversations regarding the \feiii \ line ratios and Chris DePree for kindly providing the 14.7 GHz FITS image used in Figure~\ref{c217}. 
The plots and analysis in this article made use of the Yorick programing
language\footnote{http://yorick.sourceforge.net/}. This article made use 
of the SIMBAD database at the CDS.  Based on observations
obtained at the Gemini Observatory (proposal ID GN-2006A-C-11), which
is operated by the Association of Universities for Research in
Astronomy, Inc., under a cooperative agreement with the NSF on behalf
of the Gemini partnership: the National Science Foundation (United
States), the Science and Technology Facilities Council (United
Kingdom), the National Research Council (Canada), CONICYT (Chile), the
Australian Research Council (Australia), CNPq (Brazil) and SECYT
(Argentina).

\begin{deluxetable}{lcccc}
\rotate
\label{lines}
\tablecaption{{\it Integrated Emission Line Ratios in K3--50A}}
\tablehead{
\colhead{Line Identification\tablenotemark{a}} &
\colhead{$\lambda$$_{\circ}$ (\AA)\tablenotemark{b}} &
\colhead{$\lambda$  (\AA)\tablenotemark{c}} &
\colhead{FWHM (\kms)} &
\colhead{Ratio to Br$\gamma$\tablenotemark{d}} 
}
\startdata
H$_2$ v$=$1--0 S(1)	         & 21218	& 21215 &  59 & 0.0076 $\pm$  0.0001 \\
$[$\ion{Fe}{3}$]$ $^3$H--$^3$G & 21457	& 21454&  71 & 0.0052 $\pm$  0.0001 \\ 
Br$\gamma$ (\ion{H}{1} 7--4) 	       & 21661	& 21657&  71 & 1.0000 $\pm$  0.0002 \\
$[$\ion{Kr}{3}$]$ $^3$P$_1$--$^3$P$_2$       & 21987 & 21983 & 57 & 0.0041 $\pm$  0.0001 \\
$[$\ion{Fe}{3}$]$ $^3$H--$^3$G               & 22184 & 22183 &  62 & 0.0164 $\pm$  0.0001 \\
H$_2$ v$=$1--0 S(0)	                          & 22233 & 22230 &  70 & 0.0034 $\pm$  0.0001 \\
$[$\ion{Fe}{3}$]$ $^3$H--$^3$G               & 22427 & 22422 &  54 & 0.0086 $\pm$  0.0001 \\
H$_2$ v$=$2--1 S(1)	                          & 22477 & 22473 &  55 & 0.0020 $\pm$  0.0001 \\
$[$\ion{Se}{4}$]$ $^2$P$_{3/2}$--$^2$P$_{1/2}$ & 22867 & 22863  & 71 & 0.0291 $\pm$  0.0003 

\enddata

\tablenotetext{a}{Atomic emission line identifications were made using the
on--line data base at NIST \citep{nist} and Peter van Hoof's web page
(http://www.pa.uky.edu/~peter/atomic/) that uses energy levels from
the NIST database. H$_2$ lines are from \citet{bd87}.}

\tablenotetext{b}{Rest wavelength, vacuum}

\tablenotetext{c}{Observed wavelength, mean blue shift is $-$3.8 \AA, or approximately 1.8 pixels.}

\tablenotetext{d}{Emission lines were extracted from a 2.5$''$ diameter
aperture near the center of field of view (see Figure~\ref{spec}).  The Br$\gamma$ line map is shown in Figure~\ref{brgmap}, but the line flux for this ratio was obtained by fitting multiple Gaussians to the \he1 blends and Br$\gamma$; see text.}

\end{deluxetable}

\begin{deluxetable}{lcccc}
\rotate
\label{helines}
\tablecaption{{\it Integrated \he1 Lines in K3--50A}}
\tablehead{
\colhead{Line Identification\tablenotemark{a}} &
\colhead{$\lambda$$_{\circ}$ (\AA)\tablenotemark{b}} &
\colhead{$\lambda$  (\AA)\tablenotemark{c}} &
\colhead{FWHM (\kms)} &
\colhead{Ratio to Br$\gamma$\tablenotemark{d}}
}
\startdata
\he1 $4s-6p$ $^3$S--$^3$P$^{\circ}$    & 20430      & 20427 & 79  & 0.0055 $\pm$  0.0001 \\
\he1 $2s-2p$ $^1$S--$^1$P$^{\circ}$      & 20587	& 20583 &  67 & 0.6778 $\pm$  0.0002 \\
\he1 $4p-7d$ $^3$P$^{\circ}$--$^3$D     & 20607	& 20603 & 89  & 0.0067 $\pm $ 0.0002 \\
\he1 $3p-4s$ $^3$P$^{\circ}$--$^3$S      & 21127	& 21122 &  70 & 0.0462 $\pm$  0.0001 \\
\he1 $3p-4s$ $^1$P$^{\circ}$--$^1$S      & 21138	& 21134 &  68 & 0.0110 $\pm$  0.0001 \\
\he1 $4p-7s$ $^3$P$^{\circ}$--$^3$S     & 21500     & 21497 & 70  & 0.0012 $\pm$ 0.0001 \\
\he1 $4d-7f$ $^3$D--$^3$F$^{\circ}$    & 21614	 & 21609 & 71 & 0.0250 $\pm$ 0.0002 \\
\he1 $4d-7f$ $^1$D--$^1$F$^{\circ}$     & 21623	& 21620 & 70 &  0.0100 $\pm$ 0.0002 \\
\he1 $4f-7g$ $^1$F$^{\circ}$--$^1$G, $^3$F$^{\circ}$--$^3$G  & 21647	&21642 &70 &0.0508 $\pm$ 0.0002 \\
\he1 $4f-7d$ $^1$F$^{\circ}$--$^1$D, $^3$F$^{\circ}$--$^3$D, Br$\gamma$?\tablenotemark{e}  & 21655  &21650 & 70 & 0.0670 $\pm$ 0.0002 \\
\he1 $4d-7p$ $^3$D--$^3$P$^{\circ}$    & 21821	& 21817 &62 &   0.0044 $\pm$ 0.0001 \\
\he1 $4p-7d$ $^1$P$^{\circ}$--$^1$D    & 21846	 & 21842 & 62 & 0.0023 $\pm$ 0.0001\\
\he1 $4p-7s$ $^1$P$^{\circ}$--$^1$S    & 22291 & 22287 & 78 & 0.0012 $\pm$ 0.0001 
\enddata

\tablenotetext{a}{Atomic emission line identifications were made using the
on--line data base at NIST \citep{nist} and Peter van Hoof's web page
(http://www.pa.uky.edu/~peter/atomic/) that uses energy levels from
the NIST database. Most of the lines have wavelengths calculated from the term differences.}

\tablenotetext{b}{Rest wavelength, vacuum}

\tablenotetext{c}{Observed wavelength, mean blue shift is $-$3.8 \AA, or approximately 1.8 pixels.}

\tablenotetext{d}{Emission lines were extracted from a 2.5$''$ diameter
aperture near the center of field of view (see Figure~\ref{spec}).  The Br$\gamma$ line map is shown in Figure~\ref{brgmap}, but the line flux for this ratio was obtained by fitting multiple Gaussians to the \he1 blends and Br$\gamma$; see text.}

\tablenotetext{e}{ Emission near the continuum peak is too strong to be due to \he1; see text and Figure~\ref{detail}.}

\end{deluxetable}


\begin{thebibliography}{dummy}

\bibitem[Afflerbach et al.(1997)]{aff97} Afflerbach, A., Churchwell, E., \& Werner, M.~W.\ 1997, \apj, 478, 190 

\bibitem[Barbosa et al.(2008)]{bar08} Barbosa, C.~L., Blum, R.~D., Conti, P.~S., Damineli, A., 
\& Figuer{\^e}do, E.\ 2008, \apjl, 678, L55 

\bibitem[Bautista \& Pradhan(1998)]{bau98} Bautista, M.~A., \& Pradhan, A.~K.\ 1998, \apj, 492, 650 

\bibitem[Benjamin et al.(1999)]{ben99} Benjamin, R.~A., 
Skillman, E.~D., \& Smits, D.~P.\ 1999, \apj, 514, 307 

\bibitem[Bik \& Thi(2004)]{bik04} Bik, A., \& Thi, W.~F.\  2004, \aap, 427, L13 



\bibitem[Black \& van Dishoeck(1987)]{bd87} Black, J.~H., \& 
van Dishoeck, E.~F.\ 1987, \apj, 322, 412 

\bibitem[Blum \& McGregor(2008)]{bm08} Blum, R.~D., \& McGregor,
  P.~J.\ 2008, \aj, 135, 1708

\bibitem[Blum \& Damineli(1999)]{bd99} Blum, R.~D., \& Damineli, A.\ 1999, \apj, 512, 237 

\bibitem[Blum et al.(2004)]{blum04} Blum, R.~D., Barbosa, C.~L., Damineli, A., Conti, P.~S., \& Ridgway, S.\ 2004, \apj, 617, 1167

\bibitem[Castelli \& Kurucz(2004)]{ck04} Castelli, F., \& Kurucz,
R.~L.\ 2004, ArXiv Astrophysics e-prints, arXiv:astro-ph/0405087

\bibitem[Churchwell(2002)]{church02} Churchwell, E.\ 2002, \araa, 40, 27

\bibitem[Colgan et al.(1991)]{col91} Colgan, S.~W.~J.,  Simpson, J.~P., Rubin, R.~H., Erickson, E.~F., Haas, M.~R., \& Wolf, J.\ 1991, \apj, 366, 172 

\bibitem[Colley \& Scott(1977)]{cs77} Colley, D., \& Scott, P.~F.\ 1977, \mnras, 181, 703 

\bibitem[DePoy \& Pogge(1994)]{dp94} DePoy, D.~L., \& Pogge, R.~W.\ 1994, \apj, 433, 725

\bibitem[De Pree et al.(1994)]{depree94} De Pree, C.~G.,
  Goss, W.~M., Palmer, P., \& Rubin, R.~H.\ 1994, \apj, 428, 670


\bibitem[Ferland et al.(1998)]{fer98} Ferland, G.~J., 
Korista, K.~T., Verner, D.~A., Ferguson, J.~W., Kingdon, J.~B., \& Verner, 
E.~M.\ 1998, \pasp, 110, 761 

\bibitem[Ferland(1999)]{fer99} Ferland, G.~J.\ 1999, \apj, 512, 247 



\bibitem[Frogel \& Persson(1974)]{fp74} Frogel, J.~A., \& Persson, S.~E.\ 1974, \apj, 192, 351 


\bibitem[Hanson et al.(2002)]{han02} Hanson, M.~M., Luhman, 
K.~L., \& Rieke, G.~H.\ 2002, \apjs, 138, 35

\bibitem[Harris(1975)]{har75} Harris, S.\ 1975, \mnras, 170, 139 


\bibitem[Hofmann et al.(2004)]{hof04} Hofmann, K.-H.,
  Balega, Y.~Y., Preibisch, T., \& Weigelt, G.\ 2004, \aap, 417, 981

\bibitem[Howard et al.(1997)]{how97} Howard, E.~M., Koerner, 
D.~W., \& Pipher, J.~L.\ 1997, \apj, 477, 738 

\bibitem[Howard et al.(1996)]{how96} Howard, E.~M., Pipher, 
J.~L., Forrest, W.~J., \& de Pree, C.~G.\ 1996, \apj, 460, 744 

\bibitem[Kohoutek(1965)]{k65} Kohoutek, L.\ 1965, Bulletin 
of the Astronomical Institutes of Czechoslovakia, 16, 221 

\bibitem[Kurtz et al.(1994)]{kcw94} Kurtz, S., Churchwell, 
E., \& Wood, D.~O.~S.\ 1994, \apjs, 91, 659 

\bibitem[Lacy et al.(2007)]{lacy07} Lacy, J.~H., et al.\ 2007, \apjl, 658, L45 

\bibitem[Lutz et al.(1993)]{lutz93} Lutz, D., Krabbe, A.,  \& Genzel, R.\ 1993, \apj, 418, 244 

\bibitem[Mart{\'{\i}}n-Hern{\'a}ndez et al.(2002)]{mart02} Mart{\'{\i}}n-Hern{\'a}ndez, N.~L., et al.\ 2002, \aap, 381, 606 

\bibitem[Martins et al.(2005)]{mar05} Martins, F., Schaerer, D., \& Hillier, D.~J.\ 2005, \aap, 436, 1049

\bibitem[McGregor et al.(2003)]{pm03} McGregor, P. J., Hart, J.,
Conroy, P. G., Pfitzner, M. L., Bloxham, G. J., Jones, D. J., Downing,
M. D., Dawson, M., Young, P., Jarnyk, M., \& van Harmelen, J. 2003,
SPIE, 4841, 1581.

\bibitem[Messineo et al.(2007)]{mess07} Messineo, M., Petr-Gotzens, M.~G., Schuller, F., Menten, K.~M., Habing, H.~J., Kissler-Patig, M., Modigliani, A., \& Reunanen, J.\ 2007, \aap, 472, 471

\bibitem[Morisset et al.(2002)]{mor02} Morisset, C., Schaerer, D., Mart{\'{\i}}n-Hern{\'a}ndez, N.~L., Peeters, E., Damour, F., Baluteau, J.-P., Cox, P., \& Roelfsema, P.\ 2002, \aap, 386, 558

\bibitem[Okamoto et al.(2001)]{oka01} Okamoto, Y.~K., Kataza, 
H., Yamashita, T., Miyata, T., \& Onaka, T.\ 2001, \apj, 553, 254 

\bibitem[Okamoto et al.(2003)]{oka03} Okamoto, Y.~K., Kataza, H., Yamashita, T., Miyata, T., Sako, S., Takubo, S., Honda,  M., \& Onaka, T.\ 2003, \apj, 584, 368

\bibitem[Persson \& Frogel(1974)]{pf74} Persson, S.~E., \& Frogel, J.~A.\ 1974, \apj, 188, 523 

\bibitem[Phillips \& Mampaso(1991)]{pm91} Phillips, J.~P., \& Mampaso,
  A.\ 1991, \aaps, 88, 189

\bibitem[Ralchenko et al.(2007)]{nist} Ralchenko, Yu., Jou, F.-C.,
Kelleher, D.E., Kramida, A.E., Musgrove, A., Reader, J., Wiese, W.L.,
\& Olsen, K. (2007). NIST Atomic Spectra Database (version 3.1.2),
[Online]. Available: http://physics.nist.gov/asd3 [2007, August
24]. National Institute of Standards and Technology, Gaithersburg, MD.

\bibitem[Roelfsema et al.(1988)]{roel88} Roelfsema, P.~R., Goss, W.~M., \& Geballe, T.~R.\ 1988, \aap, 207, 132

\bibitem[S{\'a}nchez(2004)]{e3d} S{\'a}nchez, S.~F.\ 2004, 
Astronomische Nachrichten, 325, 167 

\bibitem[Shields(1993)]{s93} Shields, J.~C.\ 1993, \apj, 419, 181 

\bibitem[Thompson et al.(1969)]{thom69} Thompson, A.~R., Colvin, R.~S., \& Hughes, M.~P.\ 1969, \apj, 158, 939 

\bibitem[Thompson et al.(2006)]{t06} Thompson, M.~A., Hatchell, J., Walsh, A.~J., MacDonald, G.~H., \& Millar, T.~J.\ 2006, \aap, 453, 1003 

\bibitem[Thronson \& Harper(1979)]{th79} Thronson, H.~A., Jr., \& Harper, D.~A.\ 1979, \apj, 230, 133 

\bibitem[Turner \& Matthews(1984)]{tm84} Turner, B.~E., \& Matthews,
  H.~E.\ 1984, \apj, 277, 164
  
\bibitem[Vacca et al.(1996)]{vac96} Vacca, W.~D., Garmany,  C.~D., \& Shull, J.~M.\ 1996, \apj, 460, 914 

\bibitem[Watson \& Hanson(1997)]{wat97} Watson, A.~M., \& Hanson, M.~M.\ 1997, \apjl, 490, L165

\bibitem[Wynn-Williams et al.(1977)]{ww77} Wynn-Williams, C.~G.,
  Matthews, K., Werner, M.~W., Becklin, E.~E., \& Neugebauer,
  G.\ 1977, \mnras, 179, 255


\bibitem[Yorke et al.(1984)]{york84} Yorke, H.~W., Tenorio-Tagle, G., \& Bodenheimer, P.\ 1984, \aap, 138, 325 

\end{thebibliography}
\end{document}